\pdfoutput=1
\documentclass[namedreferences]{SolarPhysics}
\usepackage[optionalrh,solaenum]{mysola} 
\usepackage{graphicx}        
\usepackage{color}           
\usepackage{url}             

\newcommand{\aap}{    {\it Astron. Astrophys.}}
\newcommand{\aaps}{   {\it Astron. Astrophys. Suppl.}}

\newcommand{\apj}{    {\it Astrophys. J.}}

\newcommand{\solphys}{{\it Solar Phys.}}

\begin{document}

\begin{article}

\begin{opening}

\title{On the Formation Height of the SDO/HMI Fe 6173\,{\AA} Doppler Signal}

\author{B.~\surname{Fleck}$^{1}$\sep
        S.~\surname{Couvidat}$^{2}$\sep
        T.~\surname{Straus}$^{3}$      
       }
\runningauthor{B. Fleck et {\it al.}}
\runningtitle{On the Formation Height of the SDO/HMI Doppler Signal}

\institute{$^{1}$ ESA Science Operations Department, c/o NASA/GSFC, Greenbelt, MD, USA
                     email: \url{bfleck@esa.nascom.nasa.gov} \\ 
              $^{2}$ W.W. Hansen Experimental Physics Laboratory, Stanford University, Stanford, CA, USA
                     email: \url{couvidat@stanford.edu} \\
$^{3}$ INAF Osservatorio Astronomico di Capodimonte, Via Moiariello 16, 80131 Napoli, Italy
                     email: \url{straus@oacn.inaf.it} \\
                                  }

\begin{abstract}
The {\it Helioseismic and Magnetic Imager} (HMI) onboard the {\it Solar Dynamics Observatory} (SDO) is designed to study oscillations and the magnetic field in the solar photosphere. It observes the full solar disk in the Fe {\sc i} absorption line at 6173\,{\AA}. We use the output of a high-resolution 3D, time-dependent, radiation-hydrodynamic  simulation based on the {\sf CO$^5$BOLD} code to calculate profiles $F(\lambda,x,y,t)$ for the Fe {\sc i }6173\,{\AA} line. The emerging profiles $F(\lambda,x,y,t)$ are multiplied by a representative set of HMI filter transmission profiles $R_i(\lambda,1 \le i \le 6)$ and filtergrams $I_i(x,y,t; 1 \le i \le 6)$ are constructed for six wavelengths. Doppler velocities $V_\mathrm{HMI}(x,y,t)$ are determined from these filtergrams using a simplified version of the HMI pipeline. The Doppler velocities are correlated with the original velocities in the simulated atmosphere. The cross-correlation peaks near 100 km, suggesting that the HMI Doppler velocity signal is formed rather low in the solar atmosphere. The same analysis is performed for the SOHO/MDI Ni {\sc i} line at 6768\,{\AA}.  The MDI Doppler signal is formed slightly higher at around 125\,km.  Taking into account the limited spatial resolution of the instruments, the apparent formation height of both the HMI and MDI Doppler signal increases by 40 to 50\,km. We also study how uncertainties in the HMI filter-transmission profiles affect the calculated velocities. 
\end{abstract}
\keywords{Sun, photosphere, oscillations, line formation}
\end{opening}

\section{Introduction}
     \label{S-Introduction} 

The {\it Helioseismic and Magnetic Imager} (HMI: \opencite{Schou2011}) onboard the {\it Solar Dynamics Observatory} 
observes the full solar disk in the Fe {\sc i} absorption line at 6173\,{\AA} to measure oscillations and the magnetic field in the solar photosphere. The oscillation measurements are used for helioseismic studies of the solar interior, the magnetic-field measurements for studies of solar activity. 

HMI supersedes the {\it Michelson Doppler Imager} (MDI: \opencite{Scherrer1995}) onboard the {\it Solar and Heliospheric Observatory} (SOHO: \opencite{Domingo1996}), which had a similar design and observing strategy. HMI offers significantly improved spatial resolution (4k $\times$ 4k detectors  {\it vs.} a 1k $\times$ 1k detector on MDI), increased image cadence (45 {\it vs.} 60\,seconds on MDI), full Stokes capabilities (MDI measured only the longitudinal component, or Stokes V), and all observables are transmitted to the ground without any further onboard processing, which allows for a much better correction of instrumental effects than would be possible with more limited on-board resources. Another significant change is the use of a different spectral line. While MDI observed the Ni {\sc i} 6768\,{\AA} line, which is also used by the GONG project, for HMI the Fe {\sc i} line at 6173\,{\AA} was selected. The reason for this switch is that the Fe {\sc i} line is magnetically more sensitive and offers better performance for vector magnetic-field measurements than the Ni {\sc i} line \cite{Norton2006}. 

The Fe {\sc i} 6173 line has been used extensively in solar magnetometry  ({\it e.g.} \opencite{Baur1980}; \opencite{Muller2000}; \opencite{Bello2008}, 2009). Its formation process has been studied by {\it e.g.} \inlinecite{Bruls1991}, \inlinecite{Norton2006}, and \inlinecite{Bello2009}. \inlinecite{Bruls1991} quote 276\,km and 238\,km for the formation height of the line core in the VAL-C atmosphere in LTE and NLTE, respectively.  \inlinecite{Norton2006} found a slightly higher value of 302\,km for the NLTE formation height of the line core in their calculations, which were also performed for the VAL-C atmosphere. \inlinecite{Bello2009} present velocity response functions, which show a broad maximum between 200 and 300 km for the line core and a maximum around 120 km for the center of gravity of the Fe 6173 line. 

\begin{figure}    
   \centerline{\includegraphics[width=1.0\textwidth]{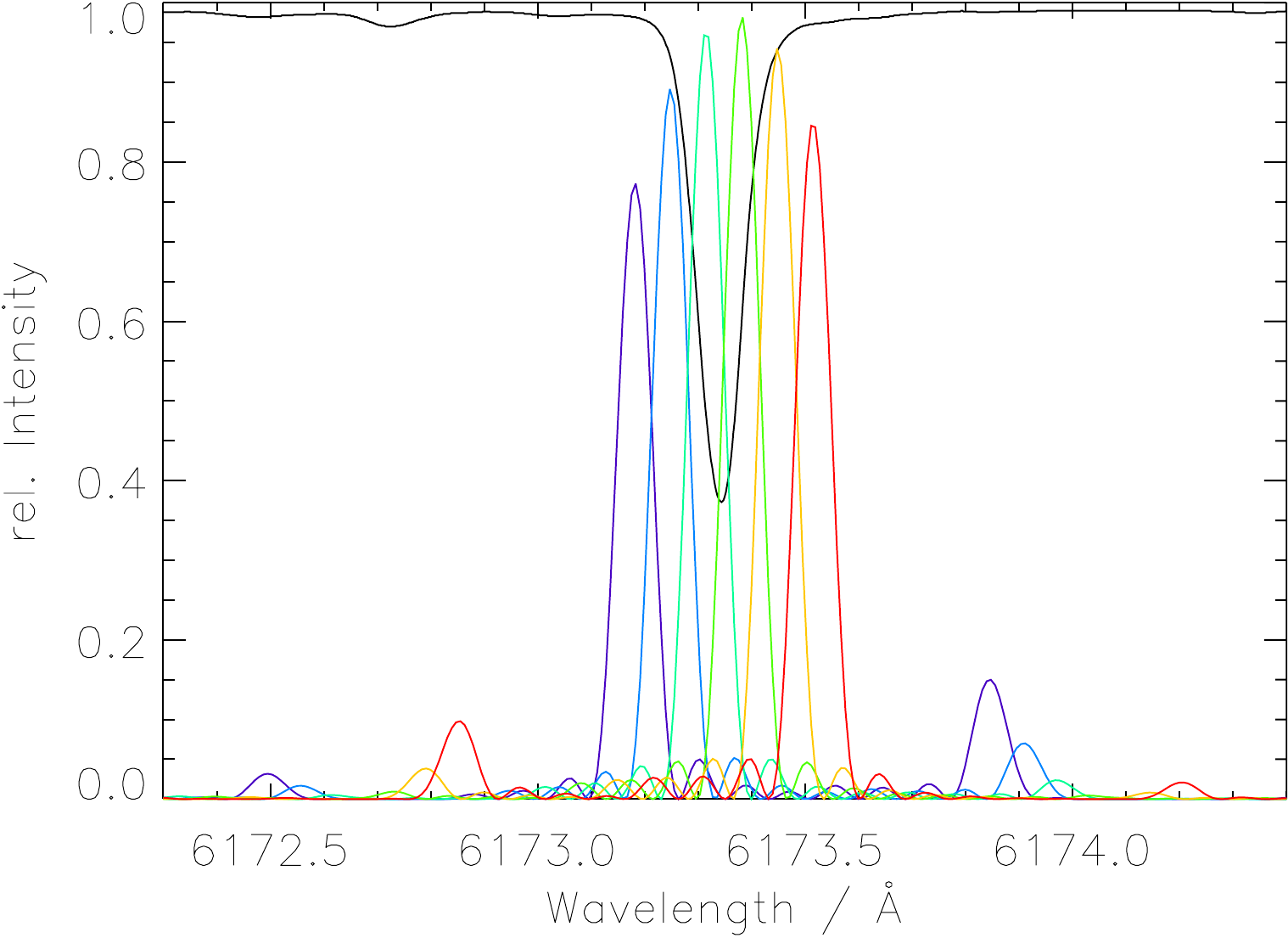}
              }
              \caption{Atlas profile of Fe 6173\,{\AA} and six representative HMI filter transmission profiles.
                      }
   \label{F-hmi-filter}
   \end{figure}

The HMI Doppler signal is not a measure of the Doppler core velocity. It is determined from a combination of six filtergrams taken around the Fe 6173\,{\AA} line. An atlas profile of the Fe 6173 line together with representative transmission profiles of the six HMI filters is shown in Figure \ref{F-hmi-filter}. The nominal HMI filter bandwith is 76\,m{\AA} and the nominal tuning range 680\,m{\AA}. This is accomplished with a set of successively narrower bandpass filters, consisting of a front window, a broadband blocking filter, a five-element Lyot filter and two Michelson interferometers. 

The aim of this study is to determine the formation height of the HMI Doppler signal, which is important for many science questions. Spectral lines are of course not formed in a narrow, static layer, but over a broad, highly variable range as the overshooting convection and {\it p} modes push the atmosphere around. The term ``formation height'' should therefore be understood as a broad average, not as a sharply defined geometrical height in a 1D, plane-parallel atmosphere.

\section{3D Model Atmosphere and Line Synthesis}
    \label{S-Model}
We study the formation of the HMI Doppler signal with the help of 3D time-dependent radiation-hydrodynamic simulations which provide a realistic description of the region of the solar atmosphere where the Fe 6173\,{\AA} line (and the Ni 6768\,{\AA} line observed by MDI) are formed. The numerical simulation of the Sun's surface layers was computed with the radiation hydrodynamics code 
{\sf CO$^5$BOLD} \cite{Freytag2002, Wedemeyer2004}. For a detailed documentation see \url{http://www.astro.uu.se/~bf/co5bold_main.html}. {\sf CO$^5$BOLD} solves the coupled non-linear equations of compressible hydrodynamics in an external gravity field together with the non-local radiative energy transport for a small fraction of the solar surface layers. A pre-tabulated equation of state, accounting for partial ionization and molecule formation, provides all necessary thermodynamic variables. The radiative transport is treated using realistic opacities adapted from a recent version of the {\sf MARCS} stellar atmosphere package \cite{Plez1992, Asplund1997, Gustafsson2003}. The simulation used in this study differs from the one used by  \inlinecite{Straus2008} in two aspects: increased spatial resolution (both horizontal and vertical) and improved treatment of radiative transport. Compared to the simulation of \inlinecite{Straus2008}, the present simulation has four times better horizontal resolution, three times better vertical resolution, and non-grey radiative transport treated in 12 frequency bands. To limit the amount of required core memory and time necessary to run the simulation, the horizontal box size was cut in half and the depth was also reduced. The simulated domain is defined by a fixed 3D Cartesian grid with $400\times 400\times 300$ cells, each of size 14\,km $\times$ 14\,km $\times$ 7.5\,km. The simulation thus covers a horizontal area of $5.6\times 5.6\,$Mm. Periodic lateral boundaries are used. The transmitting upper boundary is located at a height of approximately 900\,km above the visible solar surface, while the open lower boundary lies 1.4\,Mm below the surface. The part of the simulation that we use has completely relaxed and covers approximately two hours of solar time with snapshots saved every ten seconds. The code was executed on a 32-core node of an IBM-SP6 supercomputer and approximately three hours were necessary to advance ten seconds of solar time. 

\begin{table}
\label{T-Model}
\caption{Atomic parameters used for the line synthesis.}
\begin{tabular}{llcc}
\hline
&& Fe 6173 & Ni 6768 \\
element parameters$^a$ & & &\\
&$\log A$ & 7.52 & 6.30\\
&$W$ & 55.837 & 58.69\\

energy level parameters & & &\\
&$E_\mathrm{L}\,$[eV] & 2.2227 & 1.826 \\
&$\chi_\mathrm{I}\,$[eV] & 7.870 &  7.635 \\
&$g_\mathrm{L}$ & 3 & 1 \\
&$g_\mathrm{U}$ & 1 & 3 \\

transition parameters$^b$ & & &\\
&$\lambda_{0}$\,[{\AA}] & 6173.3352 & 6767.770 \\
&$\log gf ^c$ & -3.067 & -2.391 \\
&$\Gamma_\mathrm{r}\,[\mathrm{rad}\,\mathrm{s}^{-1}]$ & $2.042 \times 10^9$ & $1.0304 \times 10^8$ \\
&$\Gamma_5\,[\mathrm{rad}\,\mathrm{s}^{-1}\,\mathrm{cm}^3]^{c,d}$ & $3.226 \times 10^{-8}$ & $8.276 \times 10^{-8}$ \\

\hline
\multicolumn{4}{p{11cm}}{
\begin{footnotesize}
$^a$ Partition functions are interpolated following \protect\inlinecite{Allen1976}; $^b$ Stark broadening is ignored; $^c$ adapted values; $^d$ defined by $\Gamma_\mathrm{vdW}=\Gamma_5\,(T/5000)^{0.4}\,N_\mathrm{Ht}$, where $N_\mathrm{Ht}$ is the total hydrogen number density.
\end{footnotesize}
}
\end{tabular}

\end{table}

A line-synthesis code, based on the assumption of local thermodynamic equilibrium (LTE) and written in IDL, was then fed with the physical parameters of a simulation snapshot to produce synthetic, two-dimensional spectra of several photospheric lines, including those of Fe {\sc i} 6173\,{\AA} and Ni {\sc i} 6768\,{\AA}. The atomic line parameters used are listed in Table 1. The oscillator strength and van der Waals coefficient were adapted in order to match the spatial mean spectrum to a solar atlas profile. Synthetic spectra were then calculated for all snapshots with this adapted parameter set.
Using the recently published value of the iron abundance by \inlinecite{Caffau2011}, we had to reduce the oscillator strength by approximately 35\% with respect to the value given in the {\sf VALD} database \cite{Kupka1999}. Our description of the van der Waals broadening uses the modified exponent of 0.4 of the temperature dependence \cite{Gomez1987}. The value of $\Gamma_5$ in Table 1 is approximately a factor two larger than one would get by translating the parameter given in \inlinecite{Barklem2000} to our reference temperature. While such modifications are not uncommon in line-synthesis calculations, it is interesting to note that they are still necessary in a 3D simulation with such high spatial resolution. To understand the exact reason for this is not the goal of this work and difficult to resolve with only a single profile fit, but both direction and amount of the corrections seem compatible with the limits of the treatment in LTE and the uncertainties of the parameters.

\begin{figure}[h]    
   \centerline{\includegraphics[width=0.85\textwidth]{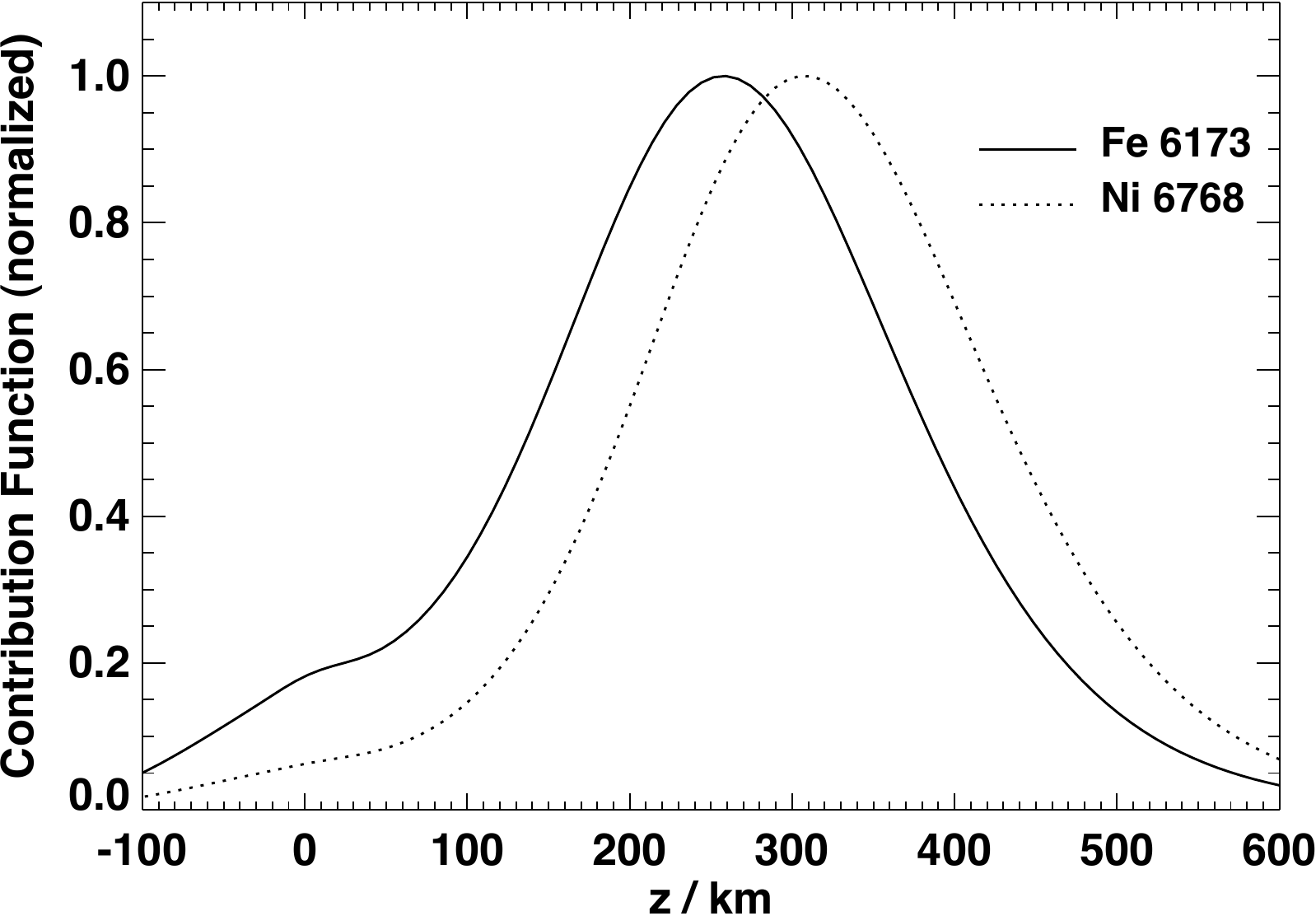}
              }
              \caption{Average intensity contribution functions of the line cores of Fe {\sc i} 6173 {\AA} and Ni {\sc i} 6768 {\AA}.                    }
   \label{F-cfs-average}
   \end{figure}

\begin{figure}[h]   
   \centerline\centerline{\hspace*{-0.03\textwidth}
               \includegraphics[width=0.515\textwidth,clip=]{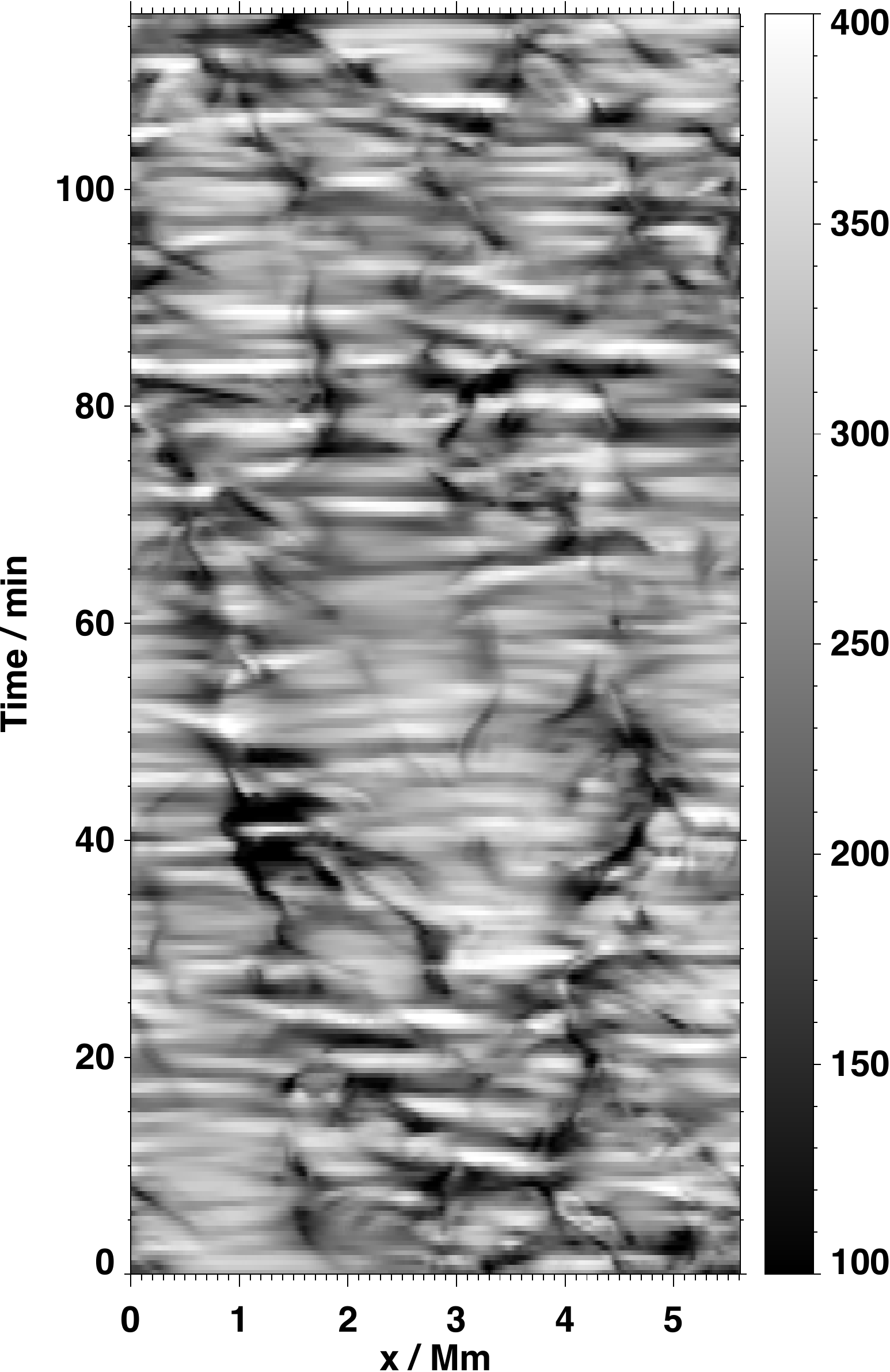}
               \hspace*{-0.03\textwidth}
               \includegraphics[width=0.515\textwidth,clip=]{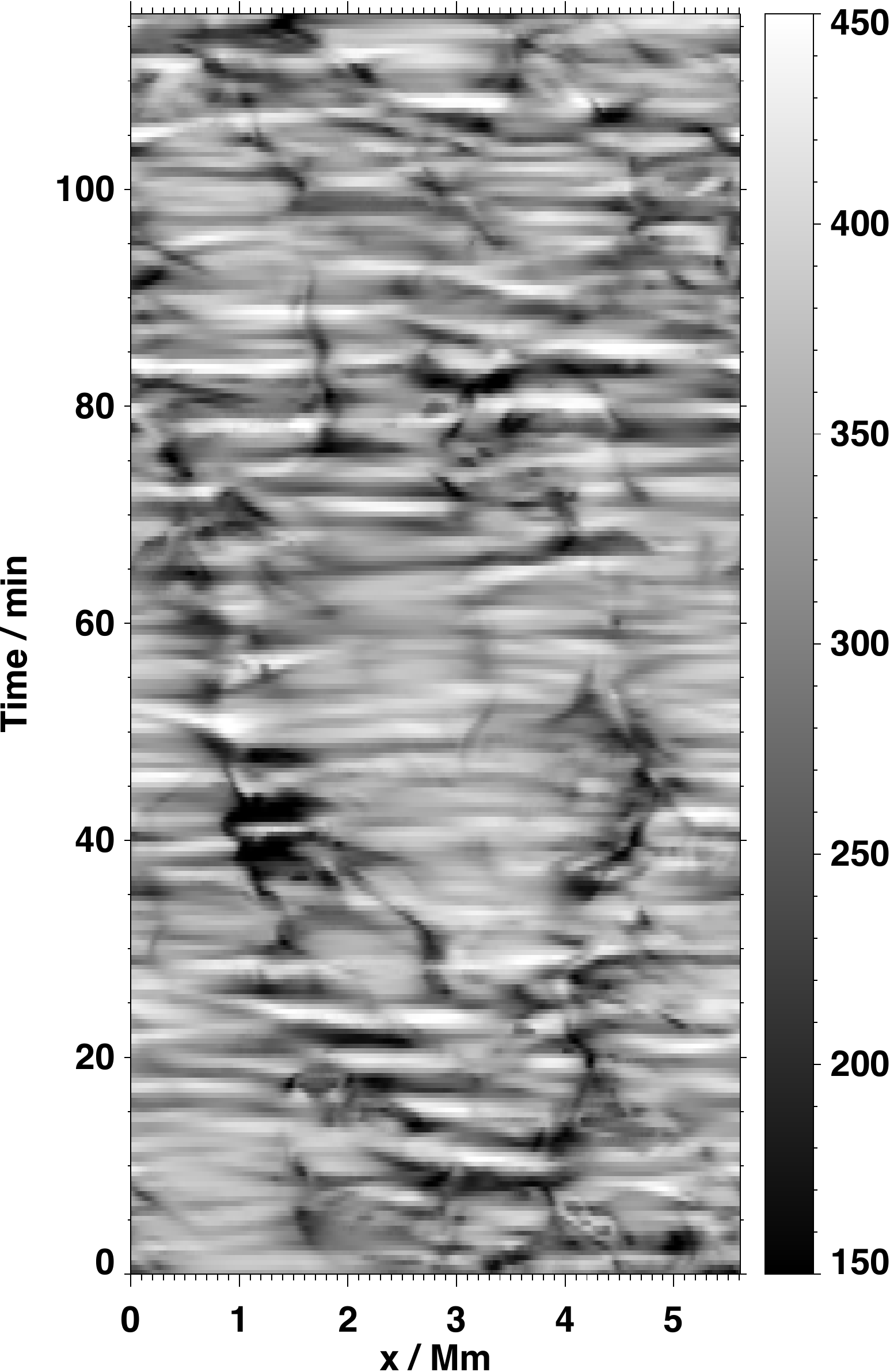}
              }

              \caption{$x$--$t$ diagrams of the height variations (center-of-gravity) of the line core intensity contribution functions of  Fe {\sc i} 6173 {\AA} (left) and Ni {\sc i} 6768 {\AA} (right). The grayscale varies between 100 and 400\,km for Fe 6173, and between 150 and 450\,km for Ni 6768.
                      }
   \label{F-cfs-x-t}
   \end{figure}

The main goal of this work is, instead, to determine the formation height of the HMI velocity signal in a realistic solar atmosphere, which is not a plane-parallel atmosphere but a highly corrugated surface.  The line contribution functions, which were stored as a by-product of the line-synthesis calculations, give a quantitative idea of the size of this effect. Figure \ref{F-cfs-average} shows the spatio-temporal averages of the intensity contribution functions of the line cores (at the wavelength of minimum intensity) of Fe {\sc i} 6173 {\AA} and Ni {\sc i} 6768 {\AA}, and Figure \ref{F-cfs-x-t} shows $x$--$t$ diagrams  of the height variations of  the center-of-gravity of the contribution functions of these two lines. Figure \ref{F-cfs-average} suggests that on average the cores of Fe 6173 and Ni 6768 are formed around 250\,km and 300\,km, respectively, in agreement with \inlinecite{Bruls1991} and \inlinecite{Norton2006}. However, as can be seen in Figure \ref{F-cfs-x-t}, there are large spatio-temporal variations of more than $\pm$150\,km ({\it i.e.} several pressure scale heights). The formation height of the core of Fe 6173 varies between approximately 100 and 400\,km, and that of Ni 6717 between approximately 150 and 450\,km. It is important to keep this in mind when talking about a ``formation height''.

\section{Formation Height of the HMI Doppler Signal}
   \label{S-HMI}

In  the following analysis we use only a 1D (slit) cut through the 3D simulations. We multiplied the line profiles $F(\lambda,x,t)$ of the Fe {\sc i} 6173\,{\AA} line calculated in the model atmosphere with a representative set of HMI filter transmission profiles $R_i(\lambda,1 \le i \le 6)$ and  summed those products to construct filtergrams $I_i(x,t; 1\le\,i\,\le 6) = \sum_\lambda F(\lambda,x,t) R_i(\lambda)$. In addition we determined the first moment (center of gravity) of the line profiles $V_\mathrm{cg}(x,t) = \int_\lambda \lambda ~ F(\lambda,x,t)\,\mathrm{d}\lambda ~ / $ $ ~ \int_\lambda F(\lambda,x,t)\,\mathrm{d}\lambda$ as well as the Doppler shift of the line core ($V_\mathrm{core}(x,t)$) by determining the position of a fourth-order polynomial fitted to  a range of $\pm$32\,m{\AA} around the minimum position. From the filtergrams $I_i(x,t; 1\le\,i\,\le 6)$ we determined Doppler velocities $V_\mathrm{HMI}(x,t)$ using a simplified version of the HMI observables-calculation pipeline, which comprises two steps: First, the phase of the first Fourier coefficient of the Fe {\sc i} line calculated from the six HMI intensities is computed, which yields an estimate of the Doppler velocity (``uncorrected velocity''): $v = k \times phase$, where $k$ is a constant. In a second step, ``corrected'' velocities are determined using a look-up table, which is based on the HMI filter-transmission profiles and an average Fe {\sc i} 6173\,{\AA} line profile (for details see \opencite{Couvidat2011}).

\begin{figure}[h]    
   \centerline\centerline{\hspace*{-0.03\textwidth}
               \includegraphics[width=0.515\textwidth,clip=]{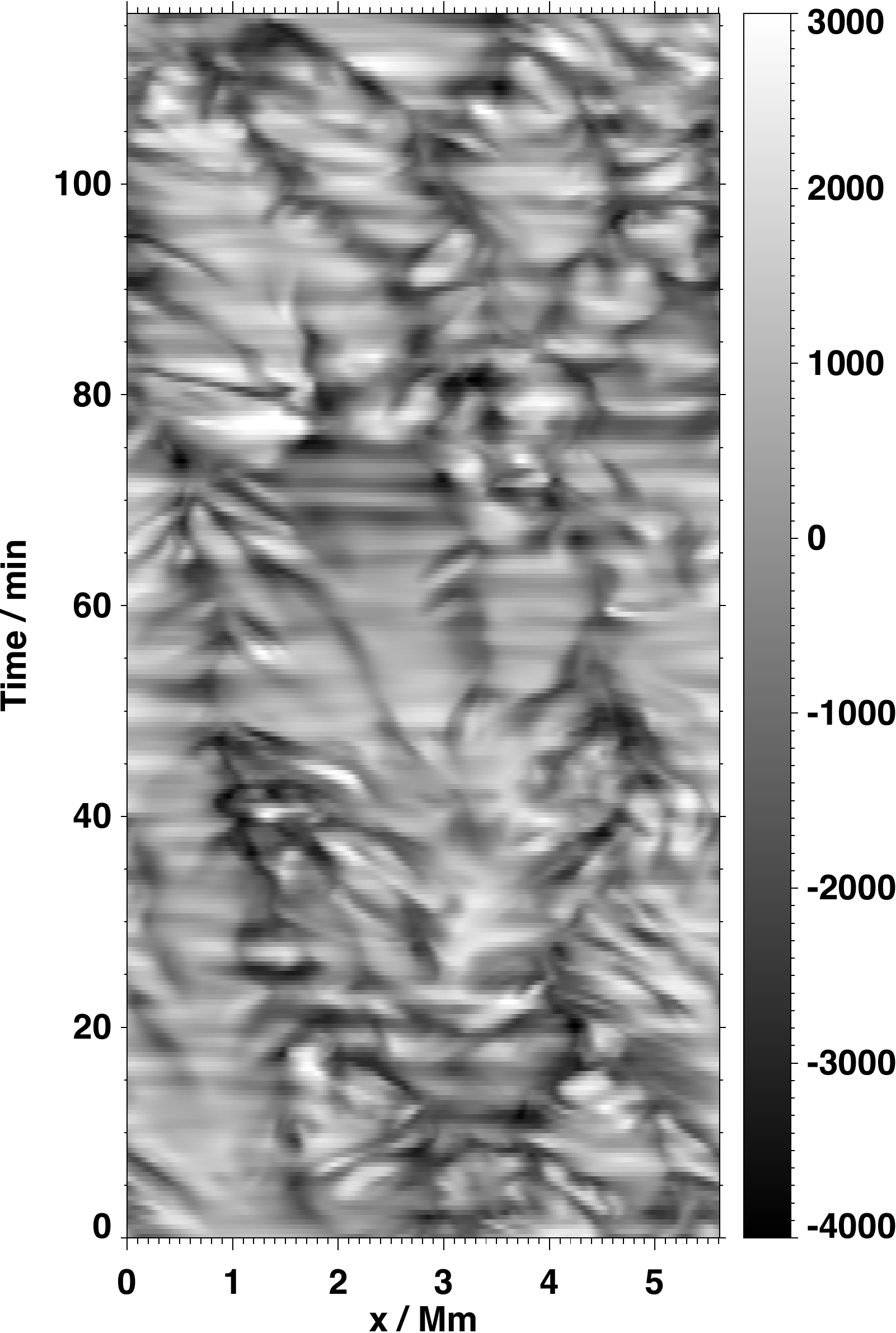}
               \hspace*{-0.03\textwidth}
               \includegraphics[width=0.515\textwidth,clip=]{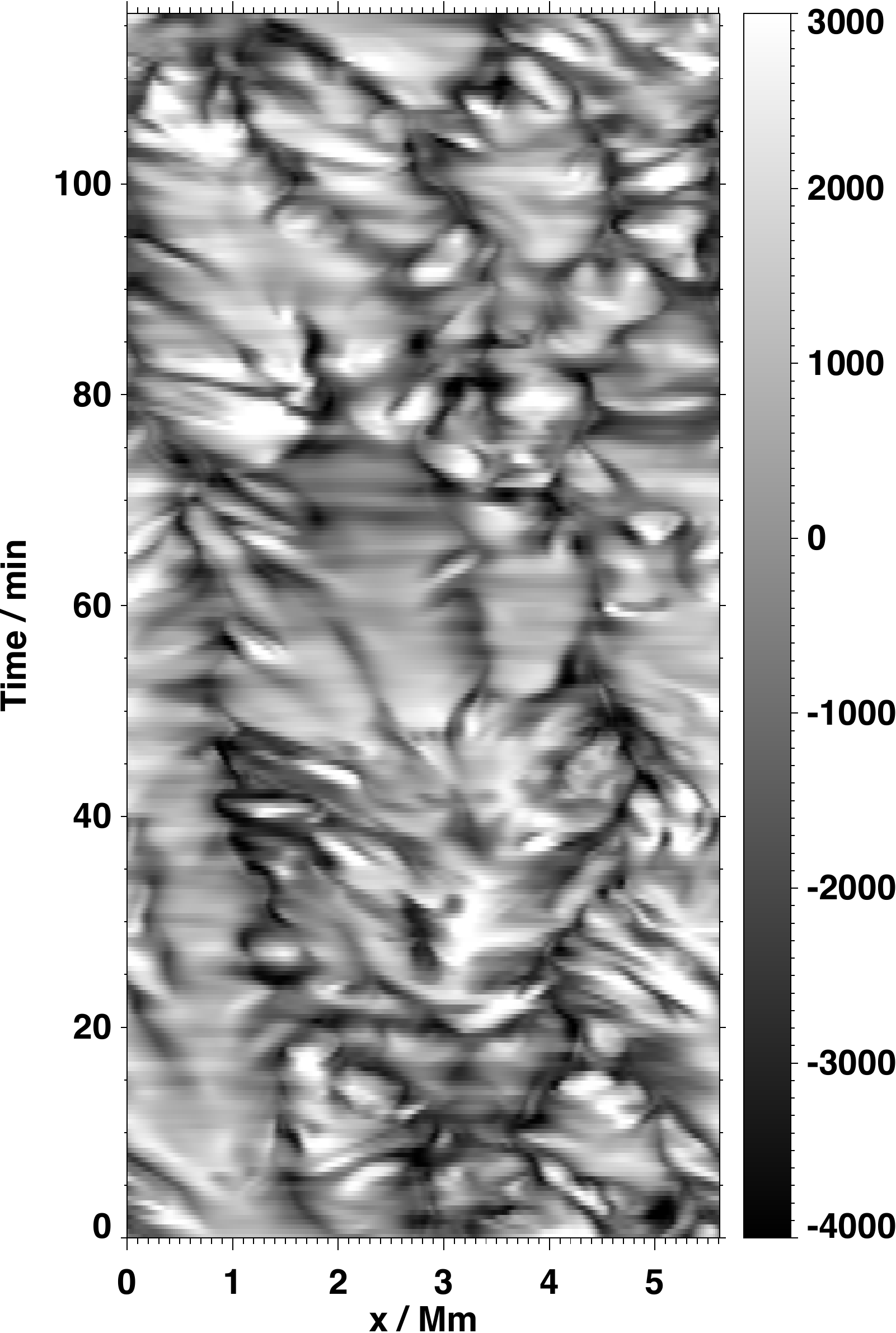}
              }

              \caption{$x$--$t$ diagrams of $V_\mathrm{HMI}(x,t)$ (left) and the actual velocities in the simulation at z = 105\,km (right). Both have the same scaling ($-4000\,\rm{m\,s^{-1}} \le V\,\le 3000\,\rm{m\,s^{-1}}$).
                      }
   \label{F-HMI-x-t-diagram}
   \end{figure}

Figure\,\ref{F-HMI-x-t-diagram} shows an $x$--$t$ diagram of the HMI Doppler velocities (left) and for comparison the actual vertical velocities in the simulation at a height of z = 105\,km (right). The corresponding $x$--$t$ diagram of the center of gravity Doppler shifts $V_\mathrm{cg}(x,t)$ (not shown here) is practically indistinguishable from that of the HMI Doppler velocities. This good correspondence is also reflected by the very high correlation (Pearson linear correlation coefficient = 0.9991) between $V_\mathrm{HMI}$ and $V_\mathrm{cg}$. 
While the HMI Doppler shifts capture the spatio-temporal variations of the actual velocities very well, the contrast is significantly (24\%) reduced: RMS($V_\mathrm{HMI}$) = 1195\ m\,s$^{-1}$ {\it vs} RMS($V_{105\,\mathrm{km}}$) = 1565\,m\,s$^{-1}$. The RMS of V$_\mathrm{cg}$ is 1254 m\,s$^{-1}$, {\it i.e.} similarly reduced. 

Given the high correlation between V$_\mathrm{HMI}$ and V$_\mathrm{cg}$, we believe that this reduction is mainly an atmospheric modulation transfer function effect, which causes all perturbations that are not coherent over the height interval where the signal is formed ({\it e.g.} high-frequency acoustic waves) to be smeared out and their amplitude reduced in the recovered Doppler signal. Another (although we believe minor) cause could be intrinsic calibration errors introduced through the correction with the look-up table. There are two issues: {\it i}) The RMS could be reduced because the slope of the look-up table is not correct and therefore, for a given phase of the first Fourier coefficient, the table returns too small a velocity. The slope could be incorrect because the average Fe {\sc i} profile used to build the table is systematically different ({\it i.e.} broader and shallower) than the actual line profiles at a given pixel. {\it ii}) The look-up table is built using an average line profile and under the assumption that the only impact of a perturbation in the atmosphere is to shift the line in wavelength, without changing its shape. This, of course, is not the case, as the thermodynamic perturbations in the atmosphere do not just cause a Doppler shift, but also affect the width, depth, and asymmetry of the individual line profiles.

\begin{figure}   
   \centerline{\includegraphics[width=0.85\textwidth]{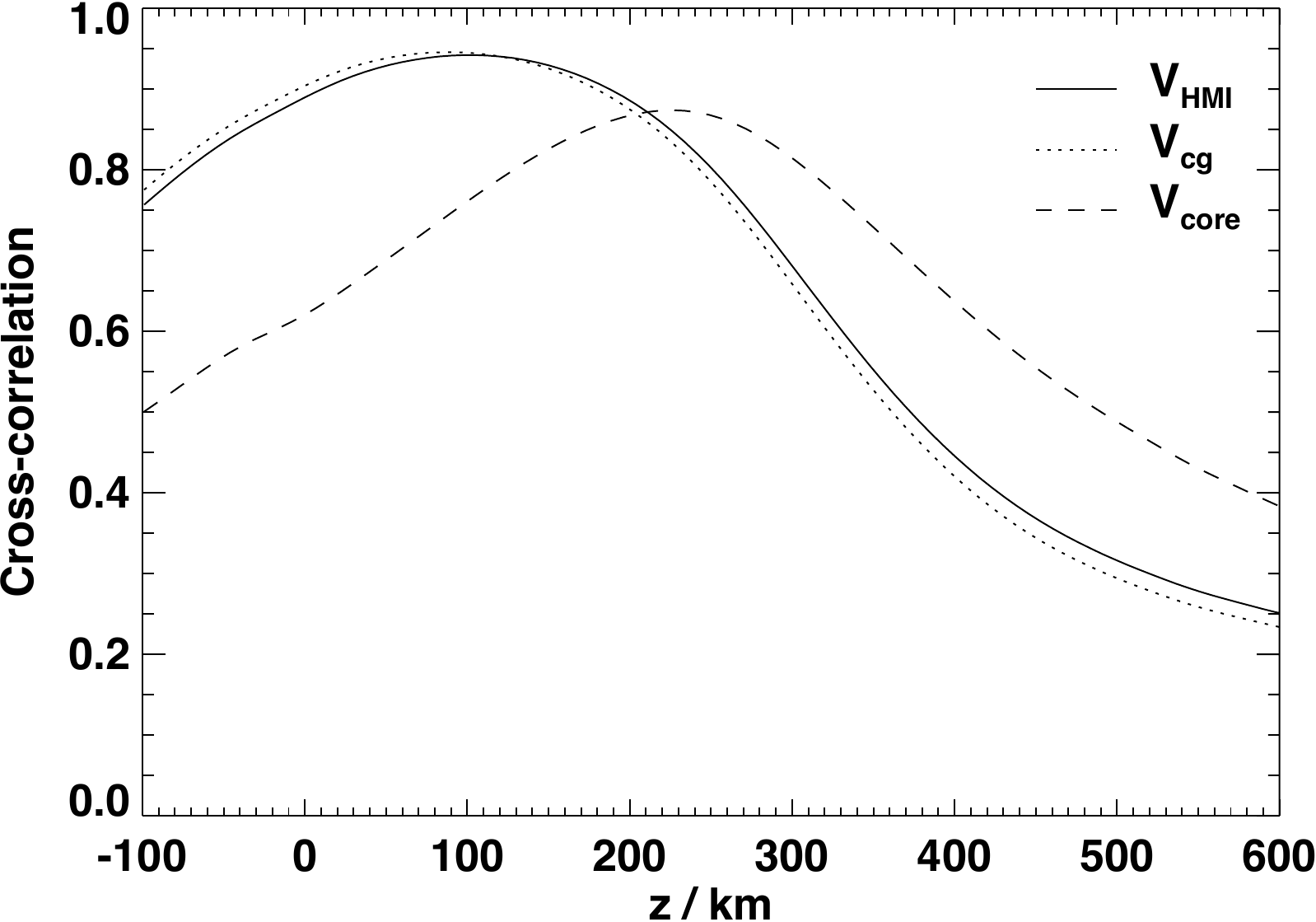}
              }
              \caption{Cross-correlations between $V_\mathrm{HMI}$,  $V_\mathrm{cg}$, $V_\mathrm{core}$ and the actual velocities $V(z)$ as a function of height in the model atmosphere.
                      }
   \label{F-HMI-CC}
   \end{figure}

To determine the formation height of the HMI Doppler signal, we calculated normalized cross-correlations $CC(z)$ between $V_\mathrm{HMI}$ and the actual velocities $V(z)$. The results are shown in Figure\,\ref{F-HMI-CC}, together with the corresponding cross-correlation functions between $V_\mathrm{cg}$ and $V(z)$, and $V_\mathrm{core}$ and $V(z)$. The cross-correlation function between $V_\mathrm{HMI}$ and $V(z)$ is very similar to that between $V_\mathrm{cg}$ and $V(z)$, confirming the excellent correspondence between $V_\mathrm{HMI}$ and $V_\mathrm{cg}$. This is perhaps not surprising, as the six HMI filters sample the whole line profile, including the upper wings reaching into the continuum. $V_\mathrm{HMI}$ peaks near 100\,km,  $V_\mathrm{cg}$ slightly below, in good agreement with the velocity-response function calculated for the center of gravity of Fe 6173\,{\AA} by \inlinecite{Bello2009}. Therefore, one can conclude that the HMI Doppler signal is a very good measure for the center of gravity of the Fe 6173\,{\AA} line, and is formed around 100\,km, {\it i.e.} rather low in the solar atmosphere.

The cross-correlation function between $V_\mathrm{core}$ and $V(z)$ peaks near 230\,km, slightly below the values given by \inlinecite{Norton2006}, but in good agreement with the NLTE value given by  \inlinecite{Bruls1991}. The height of the maximum of the cross-correlation function between $V_\mathrm{core}$ and $V(z)$ is sensitive to the wavelength range used for the polynomial fit. Extending that range from $\pm32$\,m{\AA} to $\pm56$\,m{\AA} reduces the height of the maximum of the cross-correlation function to below 200\,km. The reason for this is that by increasing the wavelength range the polynomials fitted extend into the line wings, which are formed deeper in the atmosphere. The highest layers in the atmosphere are probed only by a rather narrow wavelength range around the line core.

\section{Effects of Calibration Errors}
   \label{S-Calibration}
\begin{figure}[h]    
   \centerline{\hspace*{0.015\textwidth}
               \includegraphics[width=0.515\textwidth,clip=]{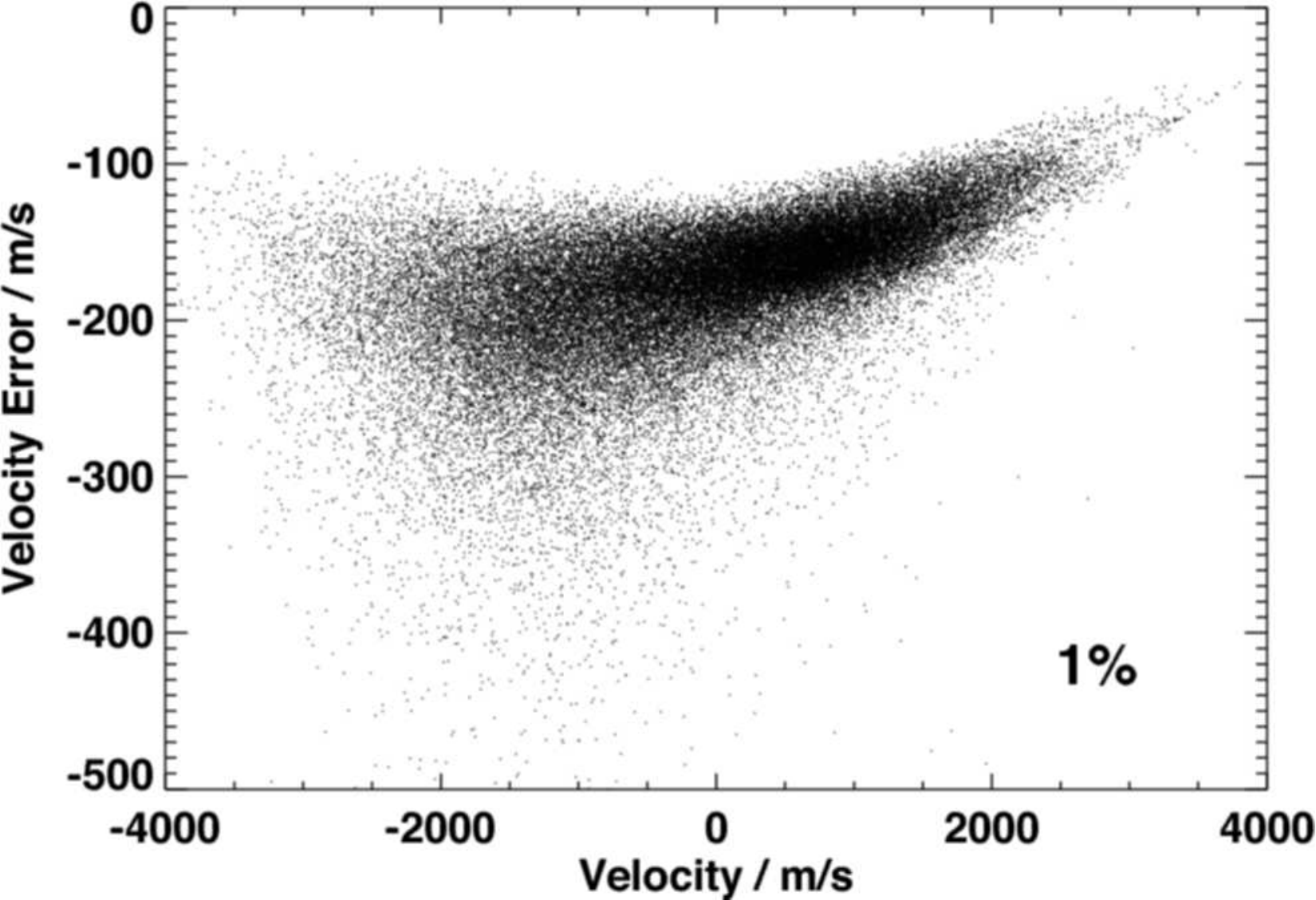}
               \hspace*{-0.03\textwidth}
               \includegraphics[width=0.515\textwidth,clip=]{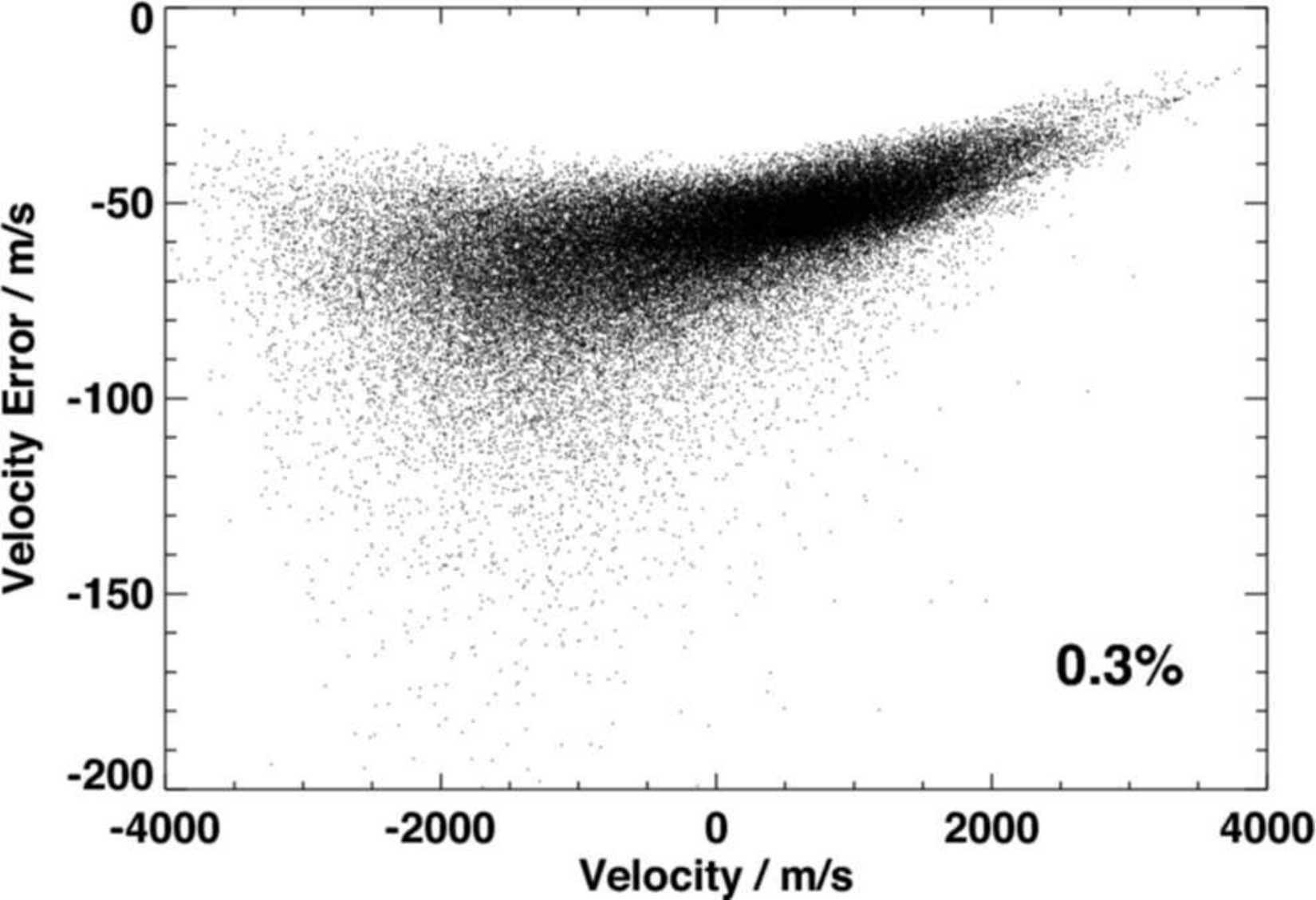}
              }
   \centerline{\hspace*{0.015\textwidth}
               \includegraphics[width=0.515\textwidth,clip=]{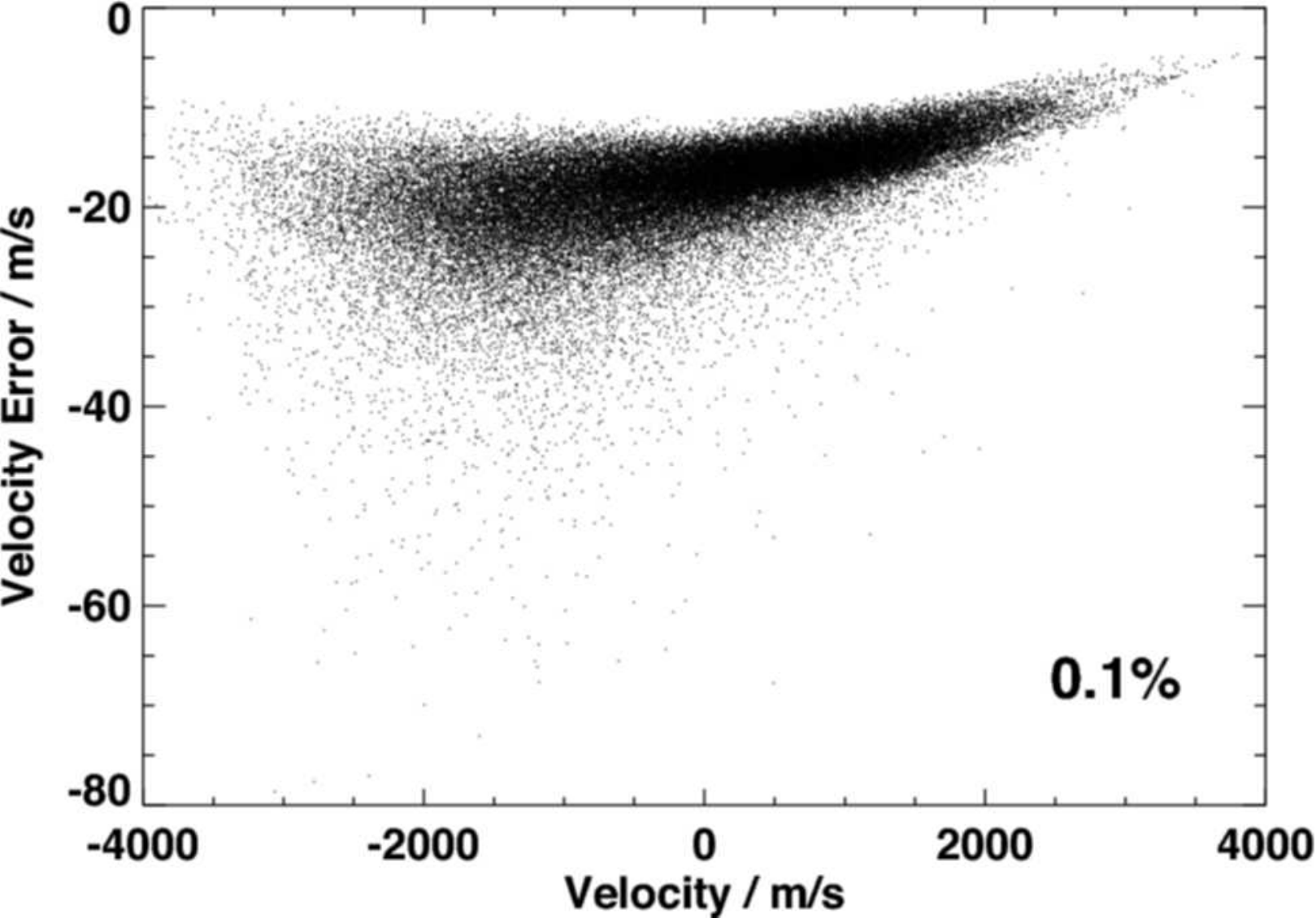}
               }
\caption{Differences between HMI Doppler shifts determined using the original filter transmission profiles and profiles that were modified by 1\% (upper left), 0.3\% (upper right), and 0.1\% (lower row). Only every fifth point is shown for clarity.
        }
   \label{F-error}
   \end{figure}

The HMI filter transmission profiles are believed to be known with an accuracy of the order of 1\% (J.~Schou, private communication). To obtain some insight into the effects of possible calibration errors on the measured velocities, we 
multiplied the blue (red) transmission profiles with a factor 1.01 (0.99), calculated new HMI intensities and from those determined new velocities using the same (unchanged) look-up tables as for the original velocities. That way we simulate unknown changes of the transmission profiles that the pipeline process cannot account for. The same is repeated for assumed transmission profile errors of 0.3\% and 0.1\%. The results are shown in Figure \ref{F-error}.

The errors are not symmetric with respect to zero velocity and show a small velocity dependence. This is mainly due to the asymmetric error introduced, although the intrinsic asymmetry of the lines and the fact that the filters are not symmetric around zero wavelength might also play a role. For the case of a 1\% (0.3\%) [0.1\%] change of the filter transmission profiles, the average offset introduced is about 170\,m\,s$^{-1}$ (60\,m\,s$^{-1}$) [17\,m\,s$^{-1}$], the RMS of the errors about 48\,m\,s$^{-1}$ (17\,m\,s$^{-1}$) [5\,m\,s$^{-1}$]. We should emphasize, though, that these are worst case figures, since the actual HMI reduction pipeline comprises a third correction step, which makes use of the orbital velocity {\sf OBS\_VR} of SDO, which is known to great accuracy. The difference,  as a function of {\sf OBS\_VR}, between  this orbital velocity and the Doppler velocity averaged over the solar disk is fitted daily by a third-order polynomial. This polynomial is then used to correct the Dopplergrams and line-of-sight magnetograms.

\begin{table}
\label{error}
\caption{Random errors (in \%) applied to the six HMI filter-response functions and resulting Doppler velocity offset and RMS errors (in m\,s$^{-1}$).}
\begin{tabular}{lccccccccc}
\hline
& 1 & 2 & 3 & 4 & 5 & 6 & &Offset & RMS  \\
&    &    &     &     &    &   &  &    [m\,s$^{-1}$]              &        [m\,s$^{-1}$]                    \\
Case 1 & 0.76 &    -0.59 &    -0.37 &   -0.025 &    0.54 &    0.96 & & -59 & 30 \\
Case 2 & -0.88 &   -0.96 &    0.41 &    -0.47 &    0.17 &     0.34 & &  -68 & 22 \\
Case 3 & -0.25 &     0.44 &    -0.35 &     0.22 &   -0.45 &    -0.46 & & 35 & 10 \\
\hline
\end{tabular}
\end{table}

\begin{figure}
   \centerline{\hspace*{0.015\textwidth}
               \includegraphics[width=0.515\textwidth,clip=]{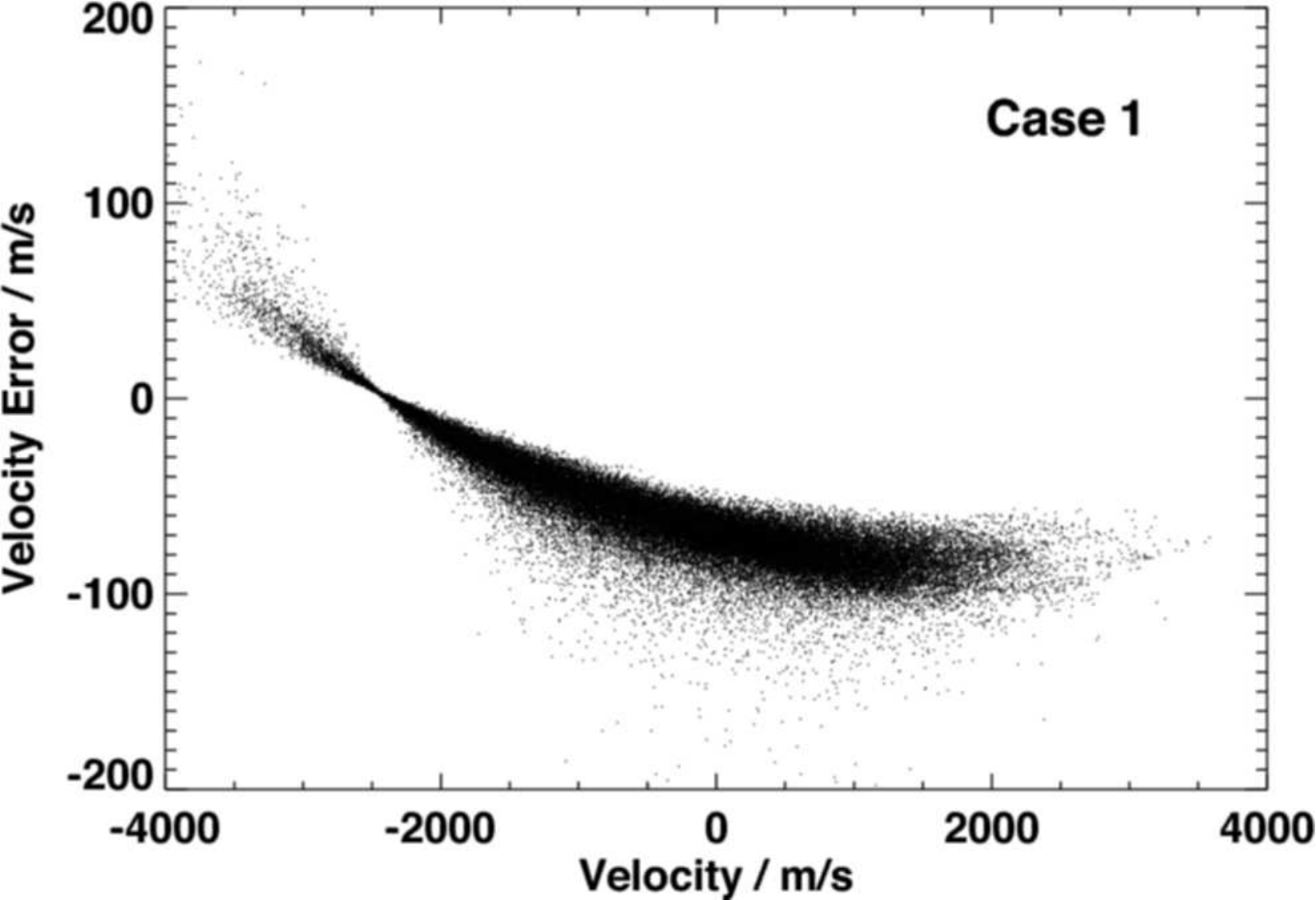}
               \hspace*{-0.03\textwidth}
               \includegraphics[width=0.515\textwidth,clip=]{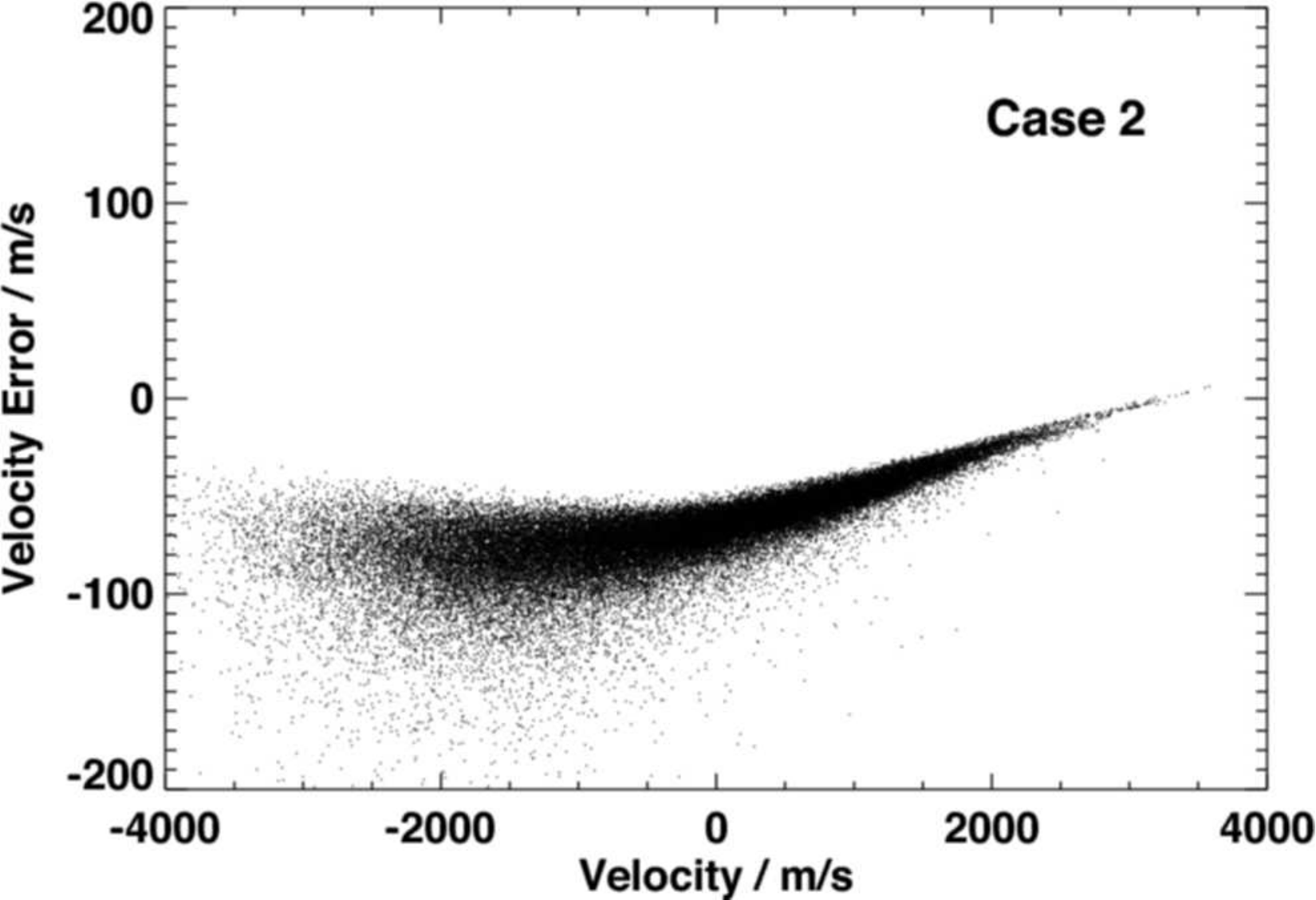}
              }
   \centerline{\hspace*{0.015\textwidth}
               \includegraphics[width=0.515\textwidth,clip=]{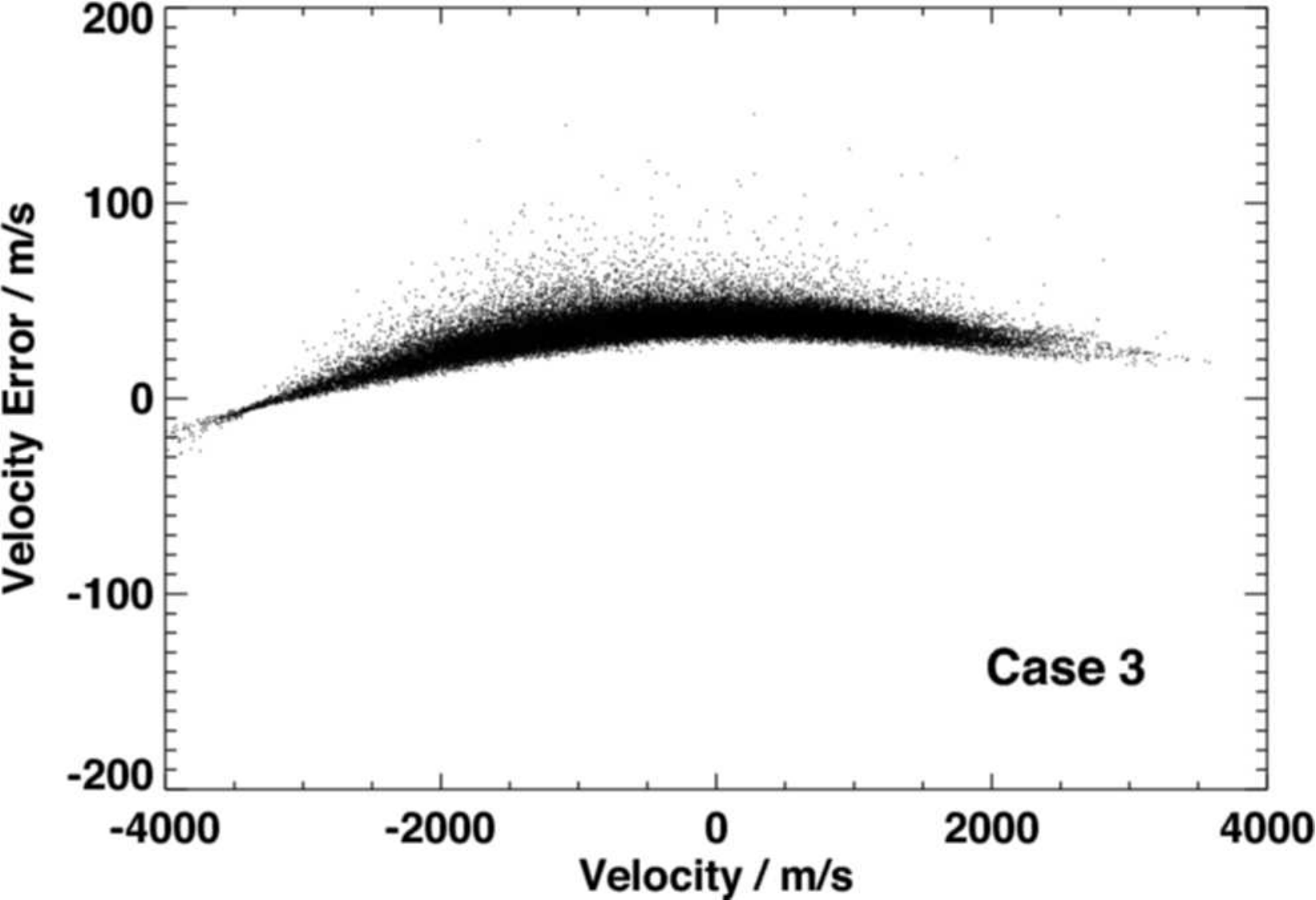}
               }
\caption{Differences between HMI Doppler shifts determined using the original filter transmission profiles and profiles that were modified by applying random errors of $\pm 1\%$ to the six HMI filters ({\it cf.} Table \ref{error} for details).
        }
   \label{F-error-random}
   \end{figure}

There are many possible sources of errors in the filter-transmission profiles. With above analysis only a single one is tested: a possible small error of the centering of the non-tunable part of the transmission profile of HMI, caused, {\it e.g.,} by a de-centering of the blocking-filter transmission profile or an error in the phase of one of the Lyot elements. While a de-centering of the non-tunable part of the transmission profile is believed to be one of the largest possible error sources, there are others that appear more like random errors. To gain some further insight, we have therefore studied the effects of randomly distributed errors by multiplying the six HMI filters with factors $(1 + \epsilon_i)$, with $\epsilon_i$ ($1 \le i \le 6$) being random numbers between -0.01 and +0.01 ({\it cf.} Table \ref{error}). The results are shown Figure \ref{F-error-random}. The offset for the three (random) cases varies between -65 and +35\,m\,s$^{-1}$ and the RMS error between 10 and 30\,m\,s$^{-1}$. The slopes of the velocity errors are very different for the three cases, and even change sign. This demonstrates that the asymmetry of the errors introduced for the tests leading to Figure \ref{F-error} is more important than the intrinsic asymmetry of the Fe line.

\section{Formation Height of the MDI Doppler Signal}
 \label{S-MDI}
\begin{figure}    
   \centerline{\includegraphics[width=1.0\textwidth]{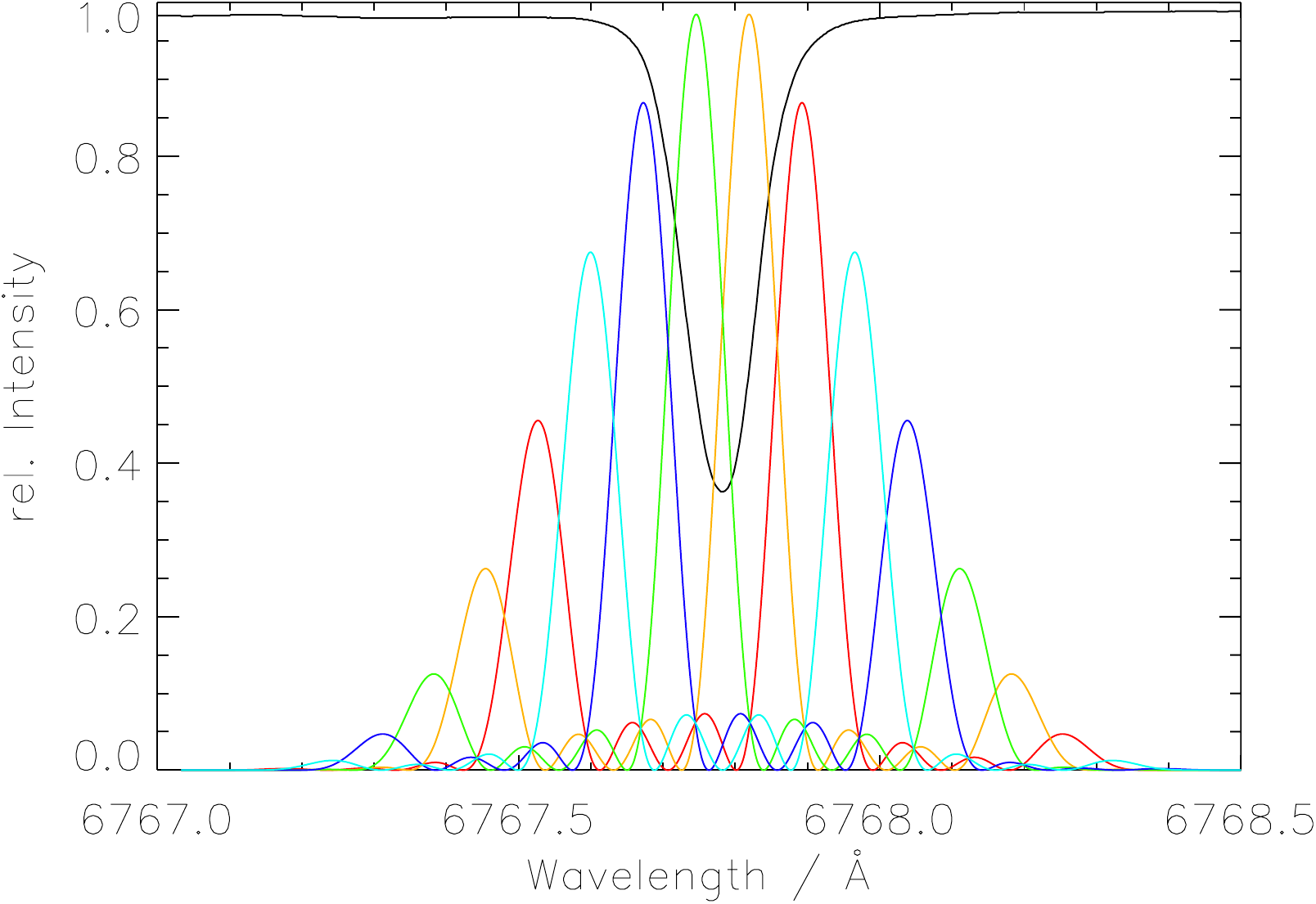}
              }
              \caption{Atlas profile of Ni 6768\,{\AA} and five representative MDI filter transmission profiles, four of which are used for the velocity determination and one (in red) to measure the continuum intensity.
                      }
   \label{F-mdi-filter}
   \end{figure}

The HMI instrument on SDO replaced the MDI instrument on SOHO, which has recorded solar oscillations for over an entire solar cycle. The HMI instrument observes in a different spectral line than MDI  and uses more and narrower filter settings. Therefore, it is reasonable to expect that the HMI and MDI Doppler signals form at different heights in the solar atmosphere. Since the HMI and MDI series might be combined for solar-cycle related studies, it is of interest to know how large this difference is.  \inlinecite{Yurchyshyn2001} adopted a formation height of 200\,km for the MDI Doppler signal based on a  study by \inlinecite{Jones1989}, \inlinecite{Straus2008} assumed 100\,km,  \inlinecite{Wachter2008} in his study on the instrumental response function of filtergraph instruments obtained 180\,km for the MDI signal for the quiet-Sun reference model of \inlinecite{Maltby1986}, and others simply use the Ni 6768\,{\AA} core formation height of 300\,km. To clarify this issue and get a better idea of the formation height of the MDI Doppler signal, we repeated the procedure as described in Section \ref{S-HMI} for the Ni {\sc i} 6768\,{\AA} line and the theoretical MDI filter-response functions, which are shown in Figure \ref{F-mdi-filter} (R.~Wachter, private communication).

\begin{figure}    
   \centerline{\includegraphics[width=0.85\textwidth]{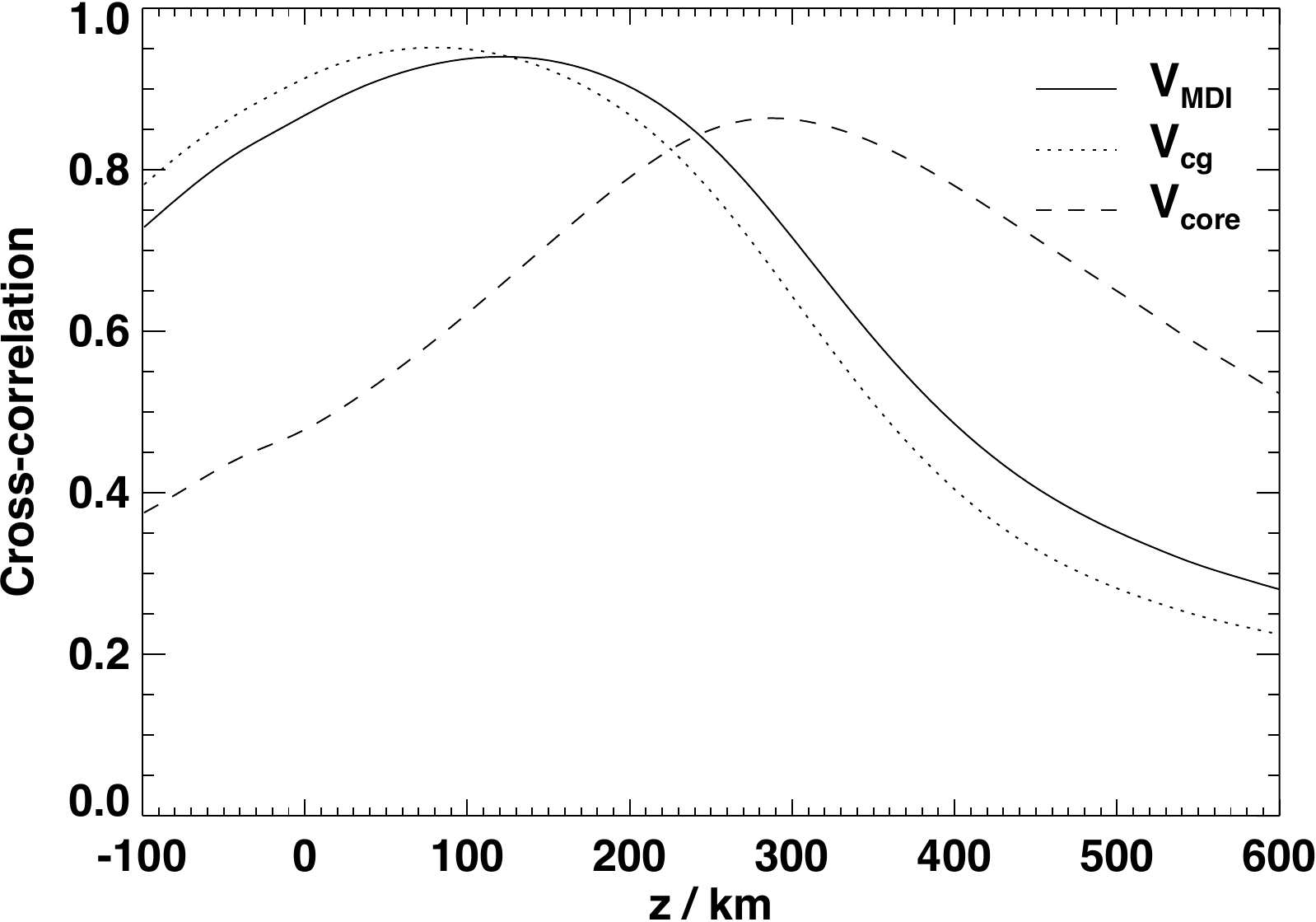}
              }
              \caption{Cross-correlations between $V_\mathrm{MDI}$,  $V_\mathrm{cg}$, $V_\mathrm{core}$ of Ni 6768\,{\AA} and the actual velocities $V(z)$ as a function of height in the model atmosphere.
                      }
   \label{F-MDI-CC}
   \end{figure}

Figure \ref{F-MDI-CC} shows the cross-correlation between the Doppler shifts $V_\mathrm{MDI}$ ``measured'' by MDI and the actual vertical velocities in the model atmosphere [$V(z)]$, together with the corresponding cross-correlation functions between $V_\mathrm{cg}$ of Ni 6768\,{\AA} and $V(z)$, and $V_\mathrm{core}$ of Ni 6768\,{\AA} and $V(z)$. Comparing this diagram to Figure \ref{F-HMI-CC} one can note several differences: {\it i}) the cross-correlation of the MDI Doppler signal ($V_\mathrm{MDI}$) is markedly different from that of the center of gravity of Ni 6768\,{\AA} ($V_\mathrm{cg}$); this is due to the fact that MDI uses only four filter positions and therefore does not sample the full profile of Ni 6768\,{\AA}; {\it ii}) the cross-correlation of $V_\mathrm{MDI}$ has its maximum at about 125\,km, {\it i.e.}~about 25\,km higher than that of $V_\mathrm{HMI}$; this is slightly  higher than the value used by \inlinecite{Straus2008}, but 55 and 75\,km lower than the values in \inlinecite{Wachter2008} and \inlinecite{Yurchyshyn2001}; and {\it iii}) the cross-correlation of $V_\mathrm{core}$ peaks at around 300\,km, significantly higher than the corresponding function of Fe 6173, and in excellent agreement with  \inlinecite{Norton2006}, who quoted 288\,km for the core formation height of the Ni line.

\section{Effects of Spatial Resolution}
With a horizontal grid size of 14\,km, the 3D simulations that we used have a much higher spatial resolution than the actual HMI (or MDI) observations. Small-scale features that are invisible in HMI observations but prominent in the simulations at certain heights might skew the results. To investigate the effects that limited spatial resolution may have on the formation height of the HMI and MDI Doppler signals, we repeated the analysis presented in Sections \ref{S-HMI}  and \ref{S-MDI} after convolving the Fe and Ni spectra in the spatial dimension with a Gaussian of 
1.5$''$ full width at half maximum (FHWM). This figure is based on a simple empirical model of the HMI point spread function (PSF) that was fitted to HMI calibration data \cite{Wachter2011}. The model includes a narrow core (described by a Gaussian) and an extended tail (accounting for scattered light). \inlinecite{Wachter2011} give a 1/e width of the central Gaussian core of 1.8 pixel, equivalent to a FWHM of approximately 1.5$''$.

\begin{figure}[h]    
   \centerline\centerline{\hspace*{-0.03\textwidth}
               \includegraphics[width=0.515\textwidth,clip=]{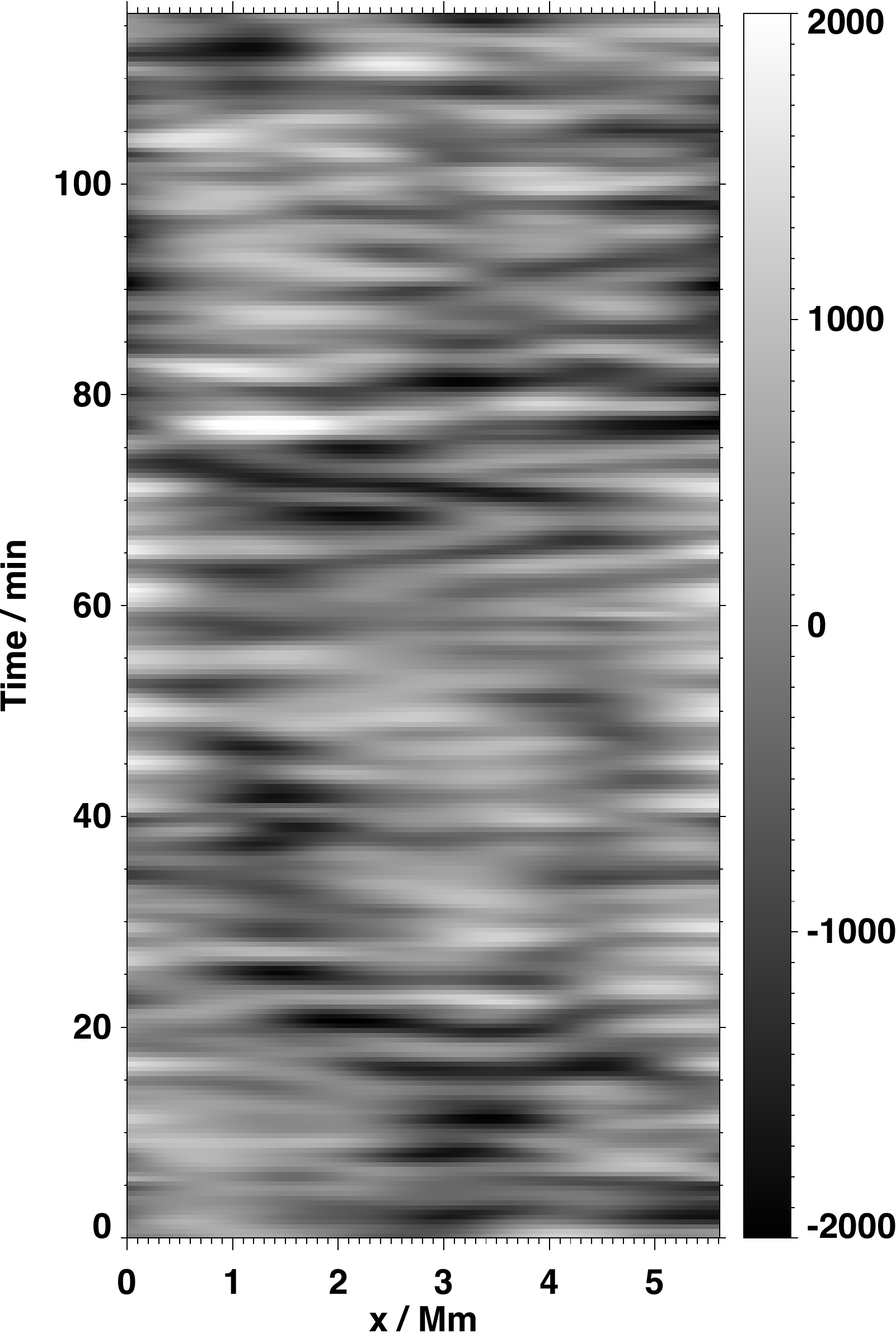}
               \hspace*{-0.03\textwidth}
               \includegraphics[width=0.515\textwidth,clip=]{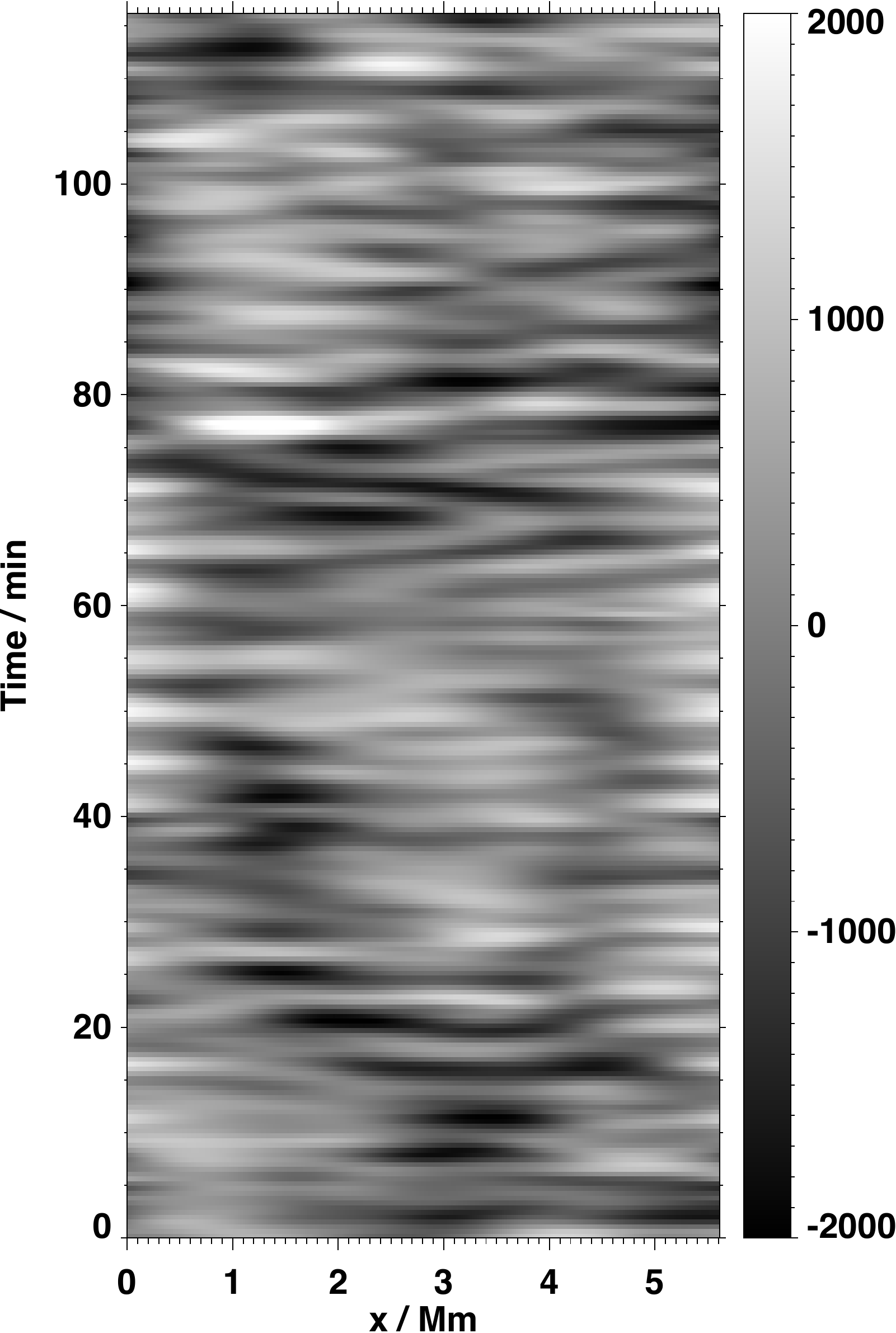}
              }

              \caption{$x$--$t$ diagrams of $V_\mathrm{HMI}(x,t)$ (left) and $V_\mathrm{MDI}(x,t)$ (right) after convolving the Fe and Ni spectra with a Gaussian of 1.5$''$ FWHM. The scaling is different from that in Figure \ref{F-HMI-x-t-diagram}. To account for the significantly reduced RMS, the range has been reduced to ($-2000\,\rm{m\,s^{-1}} \le {\it V}\,\le 2000\,\rm{m\,s^{-1}}$).
                      }
   \label{F-smeared1}
   \end{figure}

\begin{figure}   
   \centerline\centerline{\hspace*{-0.03\textwidth}
               \includegraphics[width=0.515\textwidth,clip=]{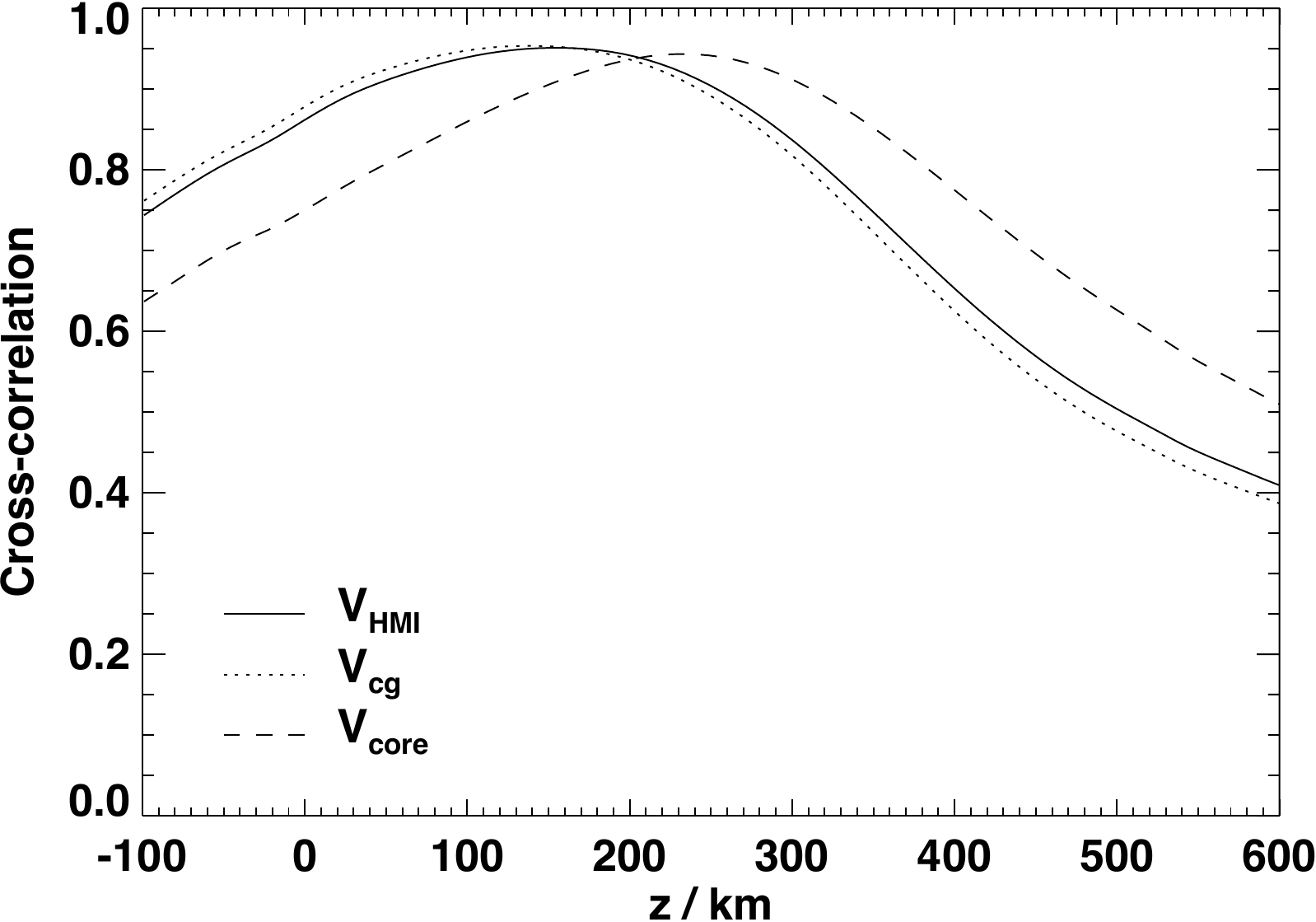}
               \hspace*{-0.03\textwidth}
               \includegraphics[width=0.515\textwidth,clip=]{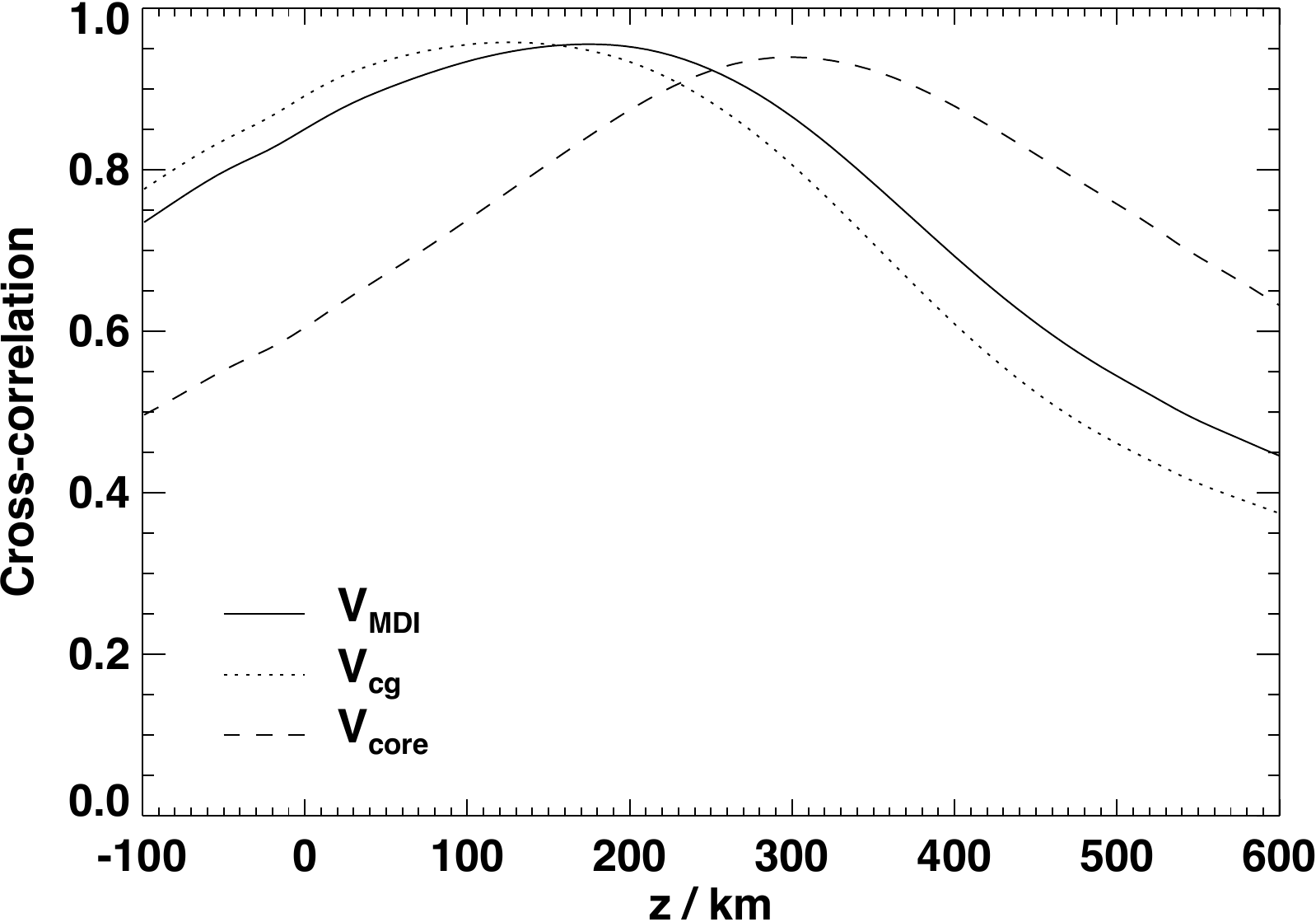}
              }

              \caption{Left: Cross-correlations between $V_\mathrm{HMI}$,  $V_\mathrm{cg}$, $V_\mathrm{core}$ and the actual velocities $V(z)$ as a function of height in the model atmosphere after convolving the Fe 6173 spectra and vertical velocities with a Gaussian of 1.5$''$ FWHM. Right: Same for the MDI Ni 6768\,{\AA} line.
                      }
   \label{F-smeared2}
   \end{figure}

 Figure \ref{F-smeared1} shows the resulting $x$--$t$ diagrams for the HMI (left) and MDI (right) Doppler signals, Figure \ref{F-smeared2} the corresponding cross-correlation functions with the spatially convolved vertical velocities in the atmosphere.

The two $x$--$t$ diagrams of the HMI and MDI Doppler signals in Figure \ref{F-smeared1} look very similar, as expected given the small difference in formation height of the two signals. Comparing them to the full-resolution $x$--$t$ diagram in Figure \ref{F-HMI-x-t-diagram}, one can notice striking differences. Most of the small-scale convective motions are lost in the signal when reduced to actual HMI resolution and the spatio-temporal pattern becomes dominated by the larger-scale five-minute {\it p} modes. As a result of this different weighting of convective motions and {\it p} modes, the apparent formation height increases by about 50\,km to 150\,km for the HMI signal and to 175\,km for the MDI signal (see the positions of the maxima of the solid lines in Figure \ref{F-smeared2}).

We should emphasize that the model PSF of \inlinecite{Wachter2011} is based on pre-launch data, not on measurements in space. Several factors typical for ground measurements (air flows, temperature changes, {\it etc.})  may have impacted the image quality. A PSF of 1.5$''$ is therefore a worst case, and it is likely that the actual HMI PSF is better. To quantify the range of the effects of limited spatial resolution, we repeated the above procedure assuming the PSF of a diffraction-limited 14-cm telescope operating at 6173\,{\AA}, which can be approximated by a Gaussian of 0.9$''$ FWHM. In that case the apparent formation height of both the HMI and MDI Doppler signal is found to be about 40\,km higher than in the full-resolution case. With the actual HMI PSF falling somewhere in between the 0.9$''$ and 1.5$''$ case, one can therefore assume a height effect due to spatial smearing of 40 to 50\,km. 

\section{Conclusions}
 \label{S-Conclusions}

We have studied the formation height of the HMI and MDI Doppler signals using 3D radiation-hydrodynamic simulations. From the synthesized Fe 6173 and Ni 6768\,{\AA} lines HMI and MDI Doppler signals were calculated using representative filter-response functions and (simplified) pipeline processes. The resulting HMI Doppler signal is nearly identical to the center-of-gravity shift of the Fe 6173\,{\AA} line (linear correlation coefficient = 0.9991). It is formed at a height of about 100\,km, about 25\,km lower than the MDI Doppler signal and significantly lower than some previous estimates. The formation height shows a small dependence on the spatial resolution and increases by 40 -- 50\,km to about 140 -- 150\, for the HMI signal and to about 165 -- 175 \,km for the MDI signal. An error of 0.3\% in the HMI filter transmission profiles (probably more realistic than a 1\% error, thanks to the {\it a-posteriori} polynomial correction applied to some observables) leads to an error of less than 20\,m\,s$^{-1}$ (RMS) in the measured HMI Doppler velocities. The current analysis is applicable only to disk center ($\mu = 1$) observations. The inclusion of center-to-limb effects would significantly complicate the foregoing analysis. A more interesting study would be to expand this analysis from HD (hydrodynamics) to MHD (magnetohydrodynamics) and investigate the magnetic-field diagnostics potential of HMI.

\begin{acks}
The simulations were carried out at CINECA (Bologna/Italy) with CPU time assigned under INAF/CINECA agreement 2008/2010. SDO is part of NASA's Living With a Star (LWS) program. HMI was designed and assembled at Stanford University and Lockheed Martin Solar and Astrophysics Laboratory. We are grateful to the SDO/HMI team for making this instrument a reality. S.C. was supported by NASA Grant NAS5-02139 (HMI). Th.S. acknowledges financial support by ASI. SOHO is a mission of international cooperation between ESA and NASA. We thank G. Severino for helpful discussions and an anonymous referee for constructive comments, which helped to improve the paper.

\end{acks}

\end{article} 


\begin{thebibliography}{}

\bibitem[\protect\citeauthoryear{Allen}{1976}]{Allen1976}
Allen, C.~W. 1976, Astrophysical Quantities, University of London, The Athlone Press, 3rd edition.

\bibitem[\protect\citeauthoryear{{Asplund {\it et al.}}}{1997}]{Asplund1997} Asplund, M., Gustafsson, B., Kiselman, D., Eriksson, K.\ 1997, \aap, \textbf{318}, 521.

\bibitem[\protect\citeauthoryear{{Barklem  {\it et al.}}}{2000}]{Barklem2000} Barklem, P.S., Piskunov, N., O'Mara, B.J.: 2000, \aaps{} \textbf{142}, 467.

\bibitem[\protect\citeauthoryear{{Baur  {\it et al.}}}{1980}]{Baur1980}
Baur,~T.G., House,~L.L., Hull,~H.K.: 1980, \solphys{} \textbf{65}, 111.

\bibitem[\protect\citeauthoryear{{Bello Gonz{\'a}lez  {\it et al.}}}{2008}]{Bello2008}
Bello Gonz{\'a}lez, N., Okunev, O., Kneer, F.: 1996, \aap{} \textbf{490}, L23.

\bibitem[\protect\citeauthoryear{{Bello Gonz{\'a}lez  {\it et al.}}}{2009}]{Bello2009}
Bello Gonz{\'a}lez, N., Yelles Chaouche, L., Okunev, O., Kneer, F.: 2009, \aap{} \textbf{494}, 1091.

\bibitem[\protect\citeauthoryear{{Bruls {\it et al.}}}{1990}]{Bruls1991}
Bruls,~J.H.M.J., Lites,~B.W., Murphy,~G.A.: 1991, In: November, L.J., (ed.), {\it Solar Polarimetry}, National Solar Observatory, Sunspot, NM, 444.

\bibitem[\protect\citeauthoryear{{Caffau {\it et al.}}}{2011}]{Caffau2011} Caffau, E., Ludwig, 
H.-G., Steffen, M., Freytag, B., Bonifacio, P.: 2011, \solphys{} \textbf{268}, 255.

\bibitem[\protect\citeauthoryear{{Couvidat {\it et al.}}}{2011}]{Couvidat2011}
Couvidat, S., Schou, J., Shine, R.A., Bush, R.I., Miles, J.W., Scherrer, P.H., Rairden, R.L.: 2011, \solphys{}, DOI: 10.1007/s11207-011-9723-8.

 \bibitem[\protect\citeauthoryear{{Domingo {\it et al.}}}{1996}]{Domingo1996}
Domingo,~V., Fleck,~B., Poland,~A.I.: 1996, \solphys{} \textbf{162}, 1.

 \bibitem[\protect\citeauthoryear{{Freytag {\it et al.}}}{2002}]{Freytag2002} Freytag, B., Steffen, 
M., Dorch, B.\ 2002, Astrono. Nach. \textbf{323}, 213.

\bibitem[\protect\citeauthoryear{{Gomez {\it et~al.}}}{1987}]{Gomez1987} 
Gomez, M.T., Severino, G., Marmolino, C., Roberti, G.: 1987, \aap{} \textbf{188}, 169.

\bibitem[\protect\citeauthoryear{{Gustafsson {\it et~al.}}}{2003}]{Gustafsson2003}
{Gustafsson}, B., {Edvardsson}, B., {Eriksson}, K., {Mizuno-Wiedner}, M.,
  {J{\o}rgensen}, U.G., {Plez}, B.: 2003, In:Hubeny, I., Mihalas, D., Werner, K. (eds)., {\it Astronomical Stellar Atmosphere Modeling}, {\bf CS-288}, Astron. Soc. of the Pacific, San Francisco, 331.


\bibitem[\protect\citeauthoryear{{Jones}}{1989}]{Jones1989}
Jones, H.P.: 1989, \solphys{} \textbf{120}, 211.

\bibitem[\protect\citeauthoryear{{Kupka {\it et~al.}}}{1999}]{Kupka1999} Kupka, F., Piskunov, N., Ryabchikova, T.A., Stempels, H.C., Weiss, W.W.: 1999, \aaps{} \textbf{138}, 119.

\bibitem[\protect\citeauthoryear{{Maltby {\it et al.}}}{1986}]{Maltby1986}
Maltby, P., Avrett, E.H., Carlsson, M., Kjeldseth-Moe, O., Kurucz, R.L., Loeser, R.: 1986 \apj{} \textbf{306}, 284.

\bibitem[\protect\citeauthoryear{{Muller {\it et al.}}}{2000}]{Muller2000}
Muller, R., Dollfus, A., Montagne, M., Moity, J., Vigneau, J.: 2000, \aap{} \textbf{359}, 373.

\bibitem[\protect\citeauthoryear{{Norton {\it et al.}}}{2006}]{Norton2006}
Norton, A.A., Pietarila Graham, J., Ulrich, R.K., Schou, J., Tomczyk, S., Liu, Y., Lites, B.W., L{'o}pez Ariste, A., Busch, R.I., Socas-Navarro, H., Scherrer, P.H.: 2006, \solphys{} \textbf{239}, 69.


\bibitem[\protect\citeauthoryear{{Plez {\it et al.}}}{1992}]{Plez1992}
{Plez}, B., {Brett}, J.M., {Nordlund}, A.: 1992, \aap{}, \textbf{256}, 551.

 \bibitem[\protect\citeauthoryear{{Scherrer {\it et al.}}}{1995}]{Scherrer1995}
Scherrer, P.H., Bogart, R.S., Busch, R.I., Hoeksema, J.T., Kosovichev, A.G., Schou, J., Rosenberg, W., Springer, L., Tarbell, T.D., Title, A., Wilfson, C.J., Zayer, I., and the MDI engineering team: 1995, \solphys{} \textbf{162}, 129.

\bibitem[\protect\citeauthoryear{{Schou {\it et al.}}}{2011}]{Schou2011}
Schou, J., Scherrer, P.H., Bush, R.I., Wachter, R., Couvidat, S., Rabello-Soares, M.C., {\it et al.}: 2011, \solphys{}, {\it in preparation}.

\bibitem[\protect\citeauthoryear{{Straus {\it et al.}}}{2008}]{Straus2008}
Straus, T., Fleck, B., Jefferies, S.M., Cauzzi, G., McIntosh, S.W., Reardon, K., Severino, G., Steffen, M.: 2008,\apj{} \textbf{681}, L125.

\bibitem[\protect\citeauthoryear{{Wachter}}{2008}]{Wachter2008}
Wachter, R.: 2008, \solphys{} \textbf{251}, 491.

\bibitem[\protect\citeauthoryear{{Wachter {\it et al.}}}{2011}]{Wachter2011}
Wachter, R., Schou, J., Rabello-Soares, M.C., Miles, J.W., Duvall, T.L., Bush, R.I.: 2011, \solphys{}, DOI: 10.1007/s11207-011-9709-6.

\bibitem[\protect\citeauthoryear{{Wedemeyer {\it et al.}}}{2004}]{Wedemeyer2004} Wedemeyer, S., Freytag, B., Steffen, M., Ludwig, H.-G., Holweger, H.\ 2004, \aap{}, \textbf{414}, 112.

\bibitem[\protect\citeauthoryear{{Yurchyshyn and Wang}}{2001}]{Yurchyshyn2001}
Yurchyshyn, V.B., Wang, H.: 2001, \solphys{} \textbf{202}, 309.




\end{thebibliography}
\end{document}